\documentclass[11pt,a4paper]{emulateapj}
\usepackage{epsfig}
\usepackage{amsmath}
\usepackage{natbib}
\usepackage{graphics}

\usepackage{color}

\begin{document}

\title{Newly-born pulsars as sources of ultrahigh energy cosmic rays}
\author{Ke Fang\altaffilmark{1}, Kumiko Kotera \altaffilmark{1,2}}
\author{Angela V. Olinto\altaffilmark{1}}
 \altaffiltext{1}{Department of Astronomy \& Astrophysics, Kavli Institute for Cosmological Physics, The
  University of Chicago, Chicago, Illinois 60637, USA.}
  \altaffiltext{2}{Theoretical Astrophysics, California Institute of Technology, 1200 E California Blvd., M/C 350-17, Pasadena, CA 91125, USA}

\begin{abstract}
Newly-born pulsars offer favorable sites for the injection of heavy nuclei, and for their further acceleration to ultrahigh energies. Once accelerated in the pulsar wind, nuclei have to escape from the surrounding supernova envelope. We examine this escape analytically and numerically, and discuss the pulsar source scenario in light of the latest ultrahigh energy cosmic ray (UHECR) data.
Our calculations show that, at early times, when protons can be accelerated to energies $E>10^{20}\,$eV, the young supernova shell tends to prevent their escape. In contrast, because of their higher charge, iron-peaked nuclei are still accelerated to the highest  observed energies at later times, when the envelope has become thin enough to allow their escape. 
Ultrahigh energy iron nuclei escape newly-born pulsars with millisecond periods and dipole magnetic fields of  $\sim10^{12-13}\,$G, embedded in core-collapse supernov\ae. Due to the production of secondary nucleons, the envelope crossing leads to a transition of composition from light to heavy elements at a few EeV, as observed by the Auger Observatory.
The escape also results in a softer spectral slope than that initially injected via unipolar induction, which allows for a good fit to the observed UHECR spectrum.
We conclude that the acceleration of iron-peaked elements in a reasonably small fraction ($\lesssim 0.01\%$) of extragalactic rotation-powered young pulsars would reproduce satisfactorily the current UHECR data. Possible signatures of this scenario are also discussed.

\end{abstract}

\section{Introduction}

The origin of the highest energy cosmic rays still remains a mystery (see \citealp{Kotera11,LSS11} for recent reviews). The  measurement of a flux suppression at the highest energies \citep{Abbasi08,Abraham10}, reminiscent of the ``GZK cut-off" \citep{G66,ZK66} produced by the interaction of particles with the cosmic microwave background (CMB) photons for propagations over intergalactic scales, has appeased the debate concerning the extragalactic provenance of UHECRs. This feature not only suggests that UHECRs would originate outside our Galaxy, but also that the sources of the highest energy particles should be located within $\sim 100~$Mpc distance, in our local Universe. However, the sources remain a mystery and results from the Auger Observatory on the arrival directions and chemical composition of UHECRs make the picture even more puzzling.

Hints of anisotropies in the sky distribution of cosmic rays above 60 EeV were reported by the Auger Observatory, but most of the anisotropy signal seems to issue from a clustering of events over a few tens of degrees around the region of Centaurus A \citep{Abreu10}. No powerful sources are observed in the direction of the highest energy events. This might be explained by strong deflections that cosmic rays could experience in presence of particularly intense extragalactic magnetic fields or if they were heavy nuclei. This absence might also find a natural explanation if the sources were transient, such as gamma-ray bursts or newly-born pulsars. The deflection in the extragalactic magnetic fields should indeed induce important time delays ($\sim 10^4$~yr for one degree deflection over 100~Mpc) between charged particles and the photons propagating in geodesics, so that the sources should already be extinguished when cosmic rays are detected on Earth. Even in this case, for proton dominated compositions and intergalactic magnetic fields of reasonable strengths, the UHECR arrival directions are expected to trace the large scale structures where the transient sources are distributed, with a possible bias \citep{Kalli10}. The precise role of extragalactic magnetic fields in UHECR propagation may be clarified in the future through extensive Faraday rotation surveys (see, e.g., \citealp{Beck07}) and indirect measurements of gamma-ray halos around blazars (e.g., \citealp{Neronov09}).

The composition measurements at the highest energies of the Auger Observatory are surprising.  \cite{Abraham:2010yv} report  a trend from a proton dominated composition at a few EeV toward an iron dominated composition at around 40 EeV (continuing up to 60 EeV, see  \citealp{AugerIcrc11}), assuming that hadronic interaction models can be extrapolated to these energies. This trend is not confirmed by the HiRes experiment \citep{Abbasi05} nor by  the preliminary data of the Telescope Array \citep{TAicrc11}, who report  light primaries in the Northern hemisphere (while Auger observes the Southern hemisphere). One may note however that both results remain consistent with those of Auger within quoted statistical and systematic errors.

From a propagation point of view, heavier nuclei are favored compared to light elements for a given energy as they can travel hundreds of megaparsecs before losing their energy by photo-disintegration processes on the cosmic backgrounds due to their lower energy per baryon (e.g., \citealp{SS99,Bertone02,Allard05,Allard08,HTS05}). Nuclei of charge $Z$ can also be in principle accelerated to an energy typically $Z$ times larger than protons in a given electromagnetic configuration. Propagation models where a heavy composition arises at the highest energies due to a combination of a low proton maximum acceleration energy (around 10~EeV) and $Z$ times higher maximum energies for heavier elements (present in a slightly higher abundance than Galactic) have been shown to reproduce the composition trends observed by Auger \citep{Allard08,Aloisio09}.  
However, these works focus on the propagation, and do not provide a plausible source for the injection of these specific compositions.
The problem of finding powerful sources that inject mainly these low abundance elements and of their escape from the acceleration site remains open.

Heavy nuclei dominated injection models are quite rare in the astrophysical literature of candidate sources. A direct injection of large proportions of heavy nuclei into an acceleration region requires either an initial metal-rich region, or an efficient nucleosynthesis in the accelerating outflow. These requirements are hardly met  by fireball-type gamma-ray bursts \citep{Lemoine02,Pruet02}. Active galactic nuclei (AGN), which are the other popular sites for UHECR acceleration models, are observed to have a solar to super-solar metallicities, but with a low proportion of nuclei heavier than nitrogen (e.g., \citealp{Groves06,Mathur09}). Young neutron stars on the other hand possess iron-rich surface and early conditions that are propitious for heavy nuclei injection.

Pulsars have been suggested as possible accelerators of cosmic-rays since their discovery \citep{Gunn69}, due to their important rotational and magnetic energy reservoirs. Galactic pulsars have been suggested as the sources of cosmic rays around the knee region up to the ankle \citep{Karakula74,Bednarek97,Bednarek02,GL02,BB04}. \cite{Blasi00} proposed that iron nuclei accelerated in the fastest spinning young neutron stars could explain the observed cosmic rays above the ankle in a Galactic source scenario. They assumed that the stripping of heavy nuclei from the surface of the star is a plausible seeding and derived a spectrum based on the spin down of young pulsars ($J \propto E^{-1}$). \cite{Arons03} studied the birth of extragalactic magnetars as the source of ultrahigh energy protons, developing the acceleration mechanism in detail and assuming that the magnetar wind disrupts the supernova envelope to allow the escape of accelerated particles.

The \cite{Blasi00} and \cite{Arons03} proposals for the origin of UHECRs were elaborated to explain the absence of the GZK cut-off  in the observed spectrum reported by AGASA \citep{Takeda98} without invoking the so-called top-down models (see, e.g., \citealp{Bhatta00}).  An increase in the exposure at the ultrahigh energies by the HiRes and Auger Observatories have shown that the UHECR spectrum is consistent with a GZK cutoff \citep{Abbasi08,Abraham10}. A decade ago, the chemical composition was also barely detectable at the highest energies while recent results suggest a puzzling trend toward heavier nuclei.  A new investigation of the pulsar scenario as UHECR sources is timely, in the light of the data that has been recently acquired.

In this paper, we examine the key mechanisms involved in the production of UHECRs by newly-born pulsars, and discuss their implications, considering the latest observational results. We focus in particular on the effects of the dense supernova envelope that surrounds the neutron star, and that accelerated particles have to traverse on their way to the interstellar medium. 
We perform detailed analytical and numerical Monte-Carlo calculations of the envelope crossing and predict the out-coming features that particles should bear after the escape. It is found that a small fraction of extragalactic rotation-powered young pulsars embedded in supernov\ae{} could satisfactorily explain the latest UHECR observations. 

The layout of this paper is the following. In Section~\ref{section:production}, we review and update the discussions related to the  production of UHE heavy nuclei in newly-born pulsars. In Section~\ref{section:escape}, we describe the supernova envelope modeling used to develop our analytical estimates and to perform our numerical simulations of the escape of UHECRs. Our main results on the escape of UHECRs from the supernov\ae{} envelopes are presented in Section~\ref{section:escape}. In Section~\ref{section:implications}  we discuss the implications of the newly-born pulsar model in view of the available UHECR observations. There, we argue how a reasonably small fraction of extragalactic {fast spinning young} pulsars embedded in supernov\ae{} could reproduce satisfactorily the current UHECR data, and discuss observable signatures that could probe the pulsar model. Our conclusions are drawn in Section~\ref{section:conclusion}.

\section{UHE heavy nuclei production \\in newly-born pulsars}\label{section:production}

In this section, we review and discuss some key points related to the production of UHE heavy nuclei in newly-born fast-spinning neutron stars. Our numerical applications focus on isolated rotation-powered pulsars of radius $R_{*,10}\equiv R_*/10\,\rm km$, angular velocity $\Omega_4\equiv\Omega/10^{4}\,{\rm s}^{-1}$, principal moment of inertia $I_{45}\equiv I/10^{45}\;\rm g\,cm^2$, and magnetic dipole moment $\mu_{30.5}\equiv \mu/10^{30.5}\,{\rm cgs}$ with
$ \mu = BR_*^3/{2} = 10^{30.5}~\mbox{cgs}\,(B/6\times10^{12}~\mbox{G})R_{*,10}^3$, with $B$ the surface dipole field strength. We show in Section~\ref{section:escape} that such parameters would enable the escape of UHE nuclei from the surrounding supernova envelope.

\subsection{Acceleration by unipolar induction}\label{subsection:unipolar}
Rapidly rotating neutron star magnetospheres are promising particle acceleration sites (see, e.g., \citealp{Shapiro83} and references therein). In the out-flowing relativistic plasma, the combination of the fast star rotation and its strong magnetic field can induce, in principle, potential differences of order $\Phi = \Omega^2\mu/c^2$. 
Provided that particles of charge $Z$ can experience a fraction $\eta$ of that potential, they can be accelerated to the energy \citep{Blasi00,Arons03}:
\begin{eqnarray}\label{eq:Eacc}
E(\Omega)=Ze\,\Phi\,\eta=3\times10^{20}\,Z_{26} \, \eta_1  \,\Omega_4^2  \,\mu_{30.5}\; \textrm{eV}
\end{eqnarray}
where $\eta_1\equiv{{\eta}/ {0.1}}$ and $Z_{26} \equiv Z/26$ for iron nuclei.

Energy losses by gravitational waves and electromagnetic radiation lead to the spin-down of the pulsar (see \citealp{Shapiro83} and references therein)\footnote{Numerical simulations of magnetized neutron star relativistic winds suggest that the spin-down rate may be  faster than obtained in the standard ``vacuum dipole" model \citep{Bucciantini06}.}, and thus to the production of particles of lower and lower energies as time goes.
Under the assumption that the Goldreich-Julian charge density \citep{Goldreich69} is entirely tapped in the outflow for acceleration, and using the expression of the pulsar spin-down rate, one can derive the energy spectrum of the accelerated particles \citep{Arons03}:
\begin{equation}\label{eq:spectrum_arons}
\frac{{\rm d} N_i}{{\rm d} E} = \frac{9}{4}\frac{c^2I}{Ze\mu} E^{-1}\left( 1+\frac{E}{E_{\rm g}}\right)^{-1}\, ,
\end{equation}
where $E_{\rm g}$ is the critical gravitational energy at which gravitational wave and electromagnetic losses are equal. The gravitational wave losses start dominating at the highest energies when the magnetic field of the star becomes $\mu\gtrsim 10^{33}\,$cgs. Magnetars are thus affected by these losses. For pulsars with milder fields that are the main concern of this paper, gravitational wave losses are negligible, and $E_{\rm g}\gg 10^{20}\,$eV.  In this case, the injected spectrum reads \citep{Blasi00}: 
\begin{equation}\label{eq:spectrum_blasi}
\frac{{\rm d} N_{\rm i}}{{\rm d} E} = 5\times10^{23} \, I_{45} (Z_{26} \, \mu_{30.5} \, E_{20})^{-1} \textrm{eV}^{-1},
\end{equation}

The spin-down time at which particles of energy $E$ can  be accelerated in the voltage drop, when gravitational wave losses are negligible, reads \citep{Arons03}:
\begin{eqnarray}\label{eq:tspin_simple}
t_{\rm spin}(E) &=& \frac{9}{8}\frac{Ic^3}{\mu^2\Omega_{\rm i}^2} \left( \frac{E_{\rm i}}{E}-1 \right)\\
&\sim& 3\times 10^{7}\, \left(\frac{3\times10^{20}\,\textrm{eV}}{E}\right) \frac{Z_{26}\eta_1I_{45}}{\mu_{30.5}}  \,{\rm s}.
\end{eqnarray}
where $E_{\rm i}$ is the maximum acceleration energy corresponding to the initial angular velocity $\Omega_{\rm i}$.  The spin-down time at which particles of energy $E$ can be accelerated does not depend on the initial rotation velocity of the neutron star $\Omega_{\rm i}$, for $E\ll E_{\rm i}$.

\subsection{Acceleration sites}\label{subsubsection:site_beyond}

Various authors have discussed particle acceleration inside the light cylinder of pulsars and magnetars (see, e.g., \citealp{Harding06} for  a review). Possible sites include the polar cap region, just above the magnetic pole of the star (e.g., \citealp{Sturrock71,Harding01,Harding02}), the ``slot gap" region along the last open field line between the polar cap and the light cylinder \citep{Arons83}, and in the outer gap region close to the light cylinder (e.g., \citealp{Cheng86, Cheng86_2,Bednarek97,Bednarek02}). Energy losses by curvature radiation are however likely to prevent the acceleration of particles to the highest energies both in the polar cap and the outer gap. \cite{Venkatesan97} and \cite{Arons03} discussed that particles accelerated in the wind region with $r\gg R_{\rm L}$ with $R_{\rm L}$ the radius of the light cylinder, do not suffer curvature radiative losses. 

In the next paragraphs, we follow the arguments of \cite{Arons03} to calculate the radius at which particle acceleration is most likely to occur. We also take into account the effects of curvature radiation of pions that was not previously considered, though it could be more constraining than the curvature radiation of photons.

Outside the light cylinder, the dipole field structure cannot be causally maintained and the field becomes mostly azimuthal, with field lines spiraling outwards \citep{Michel91}. In regions of the wind where the rest mass density is not dominated by electron and positron pairs, the plasma can be considered as force-free. In such regions, and for the case of aligned rotators, \cite{Contopoulos02} calculated that charged particles flow out with a motion along the (nearly azimuthal) magnetic field lines that becomes negligible when $r\gg r_{\rm min,lin}=\gamma_{\rm L}R_{\rm L}$. The intial Lorentz factor of the particles entering the wind, $\gamma_{\rm L}$, can take values between $10-10^3$ depending on the magnetospheric parameters. Beyond $r\gg r_{\rm min,lin}$, particles flow out nearly radially (they ``surf-ride" the fields) and the wind acts like a linear accelerator: the Lorentz factor of the out-flowing plasma increases linearly as $\gamma_{\rm w}\sim r/R_{\rm L}$. 

\cite{Arons03} extended the work of \cite{Contopoulos02} to oblique rotators and to regions in the wind where magnetic dissipation occurs (i.e., in non force-free regimes), for $r>r_{\rm diss} \sim 2\,\kappa_\pm\,R_{\rm L}$. Here $\kappa_\pm$ is the ratio between the number density of heavy ions (that we assume equal to the Goldreich-Julian density) and of electron-positron pairs. Calculations of pair creation in ultra-magnetized neutron stars suggest $\kappa_\pm\sim 10-100$ \citep{Baring01}. \cite{Arons03} discussed that surf-riding acceleration can still occur in these more general cases. He argues further that magnetic dissipation via Alfv\'en wave emission beyond $r_{\rm diss}$ would lead to an even more efficient surf-riding process, the waves acting as strong pondermotive forces on the ions. The Lorentz factor of the ions (of mass $m_{\rm i}$) would then reach values as high as $\gamma_{\rm i} = Ze\eta\Phi/(m_{\rm i}c^2) > \gamma_{\rm w}$ for $r>r_{\rm diss}$.
The results obtained for the unipolar induction toy-model described in Section~\ref{subsection:unipolar} can then be applied.

The curvature radius of a surf-riding ion at distance $r\gg r_{\rm min,lin}$ reads \citep{Arons03}: $\rho_{\rm c}=2\rho_{\rm l}\gamma_{\rm w}^2$, where $\rho_{\rm l}\sim \eta r$ is the Larmor radius of the particle.\footnote{The complete expression of the curvature radiation given by \cite{Arons03} is $\rho_{\rm c}=2\rho_{\rm l}\gamma_{\rm w}^2/\cos({\bf \Omega},{\bf \mu})$. The angle between the rotation axis and the magnetic dipole moment needs to satisfy $({\bf \Omega},\mu)<90^\circ$ to avoid curvature radiations. In such a configuration, one can expect an outflow of ions to form from the polar cap to the rotational equator, along the last closed field lines (the so-called ``return current", \citealp{Goldreich69,Michel75,Contopoulos99}). In the model of \cite{Arons03}, it is specifically this current of ions that is tapped into the wind for acceleration. }
One can calculate that, to avoid photon curvature radiation losses, the acceleration of particles at $E_{21}\equiv E/10^{21}\,{\rm eV}$ needs to take place at radius greater than:
\begin{eqnarray}\label{eq:rcmin}
r_{\rm min, c}&=&E^{1/2}\left(\frac{Z}{A^4} \frac{e^2}{6m_p^4c^4}\frac{1}{\eta \Omega^4} \right)^{1/6}\\
&\sim& 6\times 10^6 \,E_{21}^{1/2} Z_{26}^{1/6}A_{56}^{-2/3}\eta_1^{-1/6}\Omega_4^{-2/3}\, {\rm cm}\,.
\end{eqnarray}

The cooling timescale for curvature radiation of pions is more constraining; it reads \citep{Herpay08a}:
\begin{equation}
t_{\rm c,\pi} = 6\times 10^{-14}\frac{E}{10^{21}~{\rm eV}}A_{56}^{-1}\frac{\,e^{0.039/\chi}}{\chi}\,{\rm s}\ ,
\end{equation}
where $\chi\equiv E^2\hbar/(\rho_{\rm c} A^2m_p^3 c^5)$. We present here only the case of charged pions $\pi^+$, as this process dominates the case of the emission of $\pi^-$ and $\pi^0$ \citep{Herpay08}. One can readily see that $\chi\sim 13E_{21}^2A_{56}^{-2}\eta_1^{-1}\Omega_4(R_{\rm L}/r)^3\ll 1$ and thus, $t_{\rm c,\pi}\gg t_{\rm acc}$, for sufficiently large $r\gg R_{\rm L}$ in the wind. Numerically, for the same parameters as in Eq.~(\ref{eq:rcmin}),  the acceleration above $E_{21}\equiv E/10^{21}\,{\rm eV}$ needs to take place at $r>r_{\rm min,c,\pi}\sim 2\times 10^7$~cm to avoid energy losses through curvature radiation of charged pions.

The radiation fields in the pulsar wind are unlikely to impact the acceleration of UHECRs. The early neutrino-driven wind should end within the Kelvin-Helmholtz timescale of about $10-100\,$s \citep{Pons99}, and the wind should then become relativistic and non radiatively dissipative. A few days after the supernova explosion, the temperature of the soft thermal photons from the surface of the neutron star drops to $T\lesssim 10^7$~K and photo-disintegration on this background radiation can also be neglected, even inside the light cylinder \citep{Protheroe98,Bednarek02}.

In the pulsar wind beyond the light cylinder, possible acceleration sites thus lie close to the equatorial plane of the star, at a distance $R_{\rm a} >r_{\rm min}\equiv\max(r_{\rm min,lin},r_{\rm min,c},r_{\rm min,c,\pi})\sim 3\times 10^{7-9}\,\Omega_{4}^{-1}$\, cm, assuming $\gamma_{\rm L}\lesssim10^3$. 
The fact that $r_{\rm min}\gtrsim r_{\rm diss}$ implies that the unipolar induction toy-model could apply, and that particles could reach ultrahigh energies within this range of distances. 

\subsection{Heavy nuclei injection}\label{subsection:injection}

One can mention three channels via which heavy ions could be seeded in the neutron star wind.
Note that scenarios of pulsar winds loaded with heavy nuclei give a satisfactory  explanation to some observations. For instance, the morphological features of the Crab Nebula could be the signature of resonant scattering of pairs of electrons and positrons by heavy nuclei \citep{Hoshino92,Gallant94}.

The classical argument that applies best in our scenario is that iron nuclei can be stripped off the neutron star surface, as has been suggested by \cite{Ruderman75} and \cite{Arons79}. Strong electric fields combined with bombardment by particles can extract ions from the polar cap regions, where the co-rotation charge is positive provided that ${\bf \Omega}\cdot {\bf B}<0$. The surface of a neutron star being composed mainly of iron-peaked elements, it is possible that heavy nuclei get injected in the wind by these means.

Heavy nuclei loading of the pulsar wind by mixing of the stellar material via Kelvin-Helmholtz instabilities or oblique shocks was also proposed \citep{Zhang03,Wang08}. This mechanism requires however that a jet goes through the stellar core, a case that is not considered in the present study. Kelvin-Helmholtz instabilities might also occur at the interface between the wind nebula and the supernova remnant \citep{Jun98,vanderSwaluw04}, but it is unlikely that the envelope in that region has a metallicity high enough to mix large amounts of heavy nuclei in the wind.

The nucleosynthesis of heavy elements by $r$-process in the neutrino-driven wind at the very early phase of the proto-magnetar formation has also been discussed by \cite{Metzger11,Metzger10}. These authors find that the production rate of nuclei with $A\gtrsim 56$ can be important during the first 1 to $\sim\,$a few $100$~s, when the electron fraction $Y_e$ could be fairly low, the wind expansion time $\tau_{\rm exp}\lesssim 10^3$~s, and the entropy $S\lesssim100\,k_{\rm b}\,$nucleon$^{-1}$, as is required for a successful $r$-process (see, e.g., \citealp{Hoffman97}). Though these results are obtained for the case of a highly magnetized proto-magnetar driving a jet (as in \citealp{Bucciantini07}), they can be applied in a non-collimated mildly magnetized wind case, as the evolution of $S$ and $\tau_{\rm exp}$ is mostly ruled by thermal ingredients (and the rotation speed) in the times considered. However, we will see in the next section that the supernova envelope at $t\sim 10-100$~s is too dense to allow the escape of particles, whatever their mass number.
At later times, as the wind cools and becomes relativistic, the neutrino heating efficiency drops, shutting off the $r$-process.
It is thus unlikely that this channel can seed heavy nuclei in the wind in our framework.

\section{UHECR escape from supernova envelopes}\label{section:escape}

Particles accelerated in the pulsar wind further need to escape from the pulsar wind nebula itself, and then from the surrounding young supernova envelope. We assume in this study that the supernova envelope is not totally disrupted by the wind, and that particles do not escape through a region punctured by a jet, like in a strongly magnetized proto-magnetar scenario discussed by \cite{Metzger11} ---see Appendix~\ref{app:alternative} for further discussions. 

The escape of accelerated ions from the magnetar wind nebula was discussed by \cite{Arons03}.
In Section~\ref{subsubsection:site_beyond}, we argued that at distances $r\gg R_{\rm L}$, the curvature radius of the ions reads: $\rho_{\rm c}\sim 2\eta r^3/R_{\rm L}\gg r$. Hence, particles are not coupled to the magnetic field lines and can escape the wind beyond $r_{\rm min}$. 

In supernova envelopes, magnetic fields are of order a few mG at most (see, e.g., \citealp{Reynolds11} for a review). The Larmor radius of the ions is thus much larger than the size of the envelope and their trajectories can be treated rectilinearly.
We give in the following section, estimates of the density profile and composition of young supernova envelopes, that we use to study the escape of UHECRs analytically and numerically.

\subsection{Supernova envelopes}\label{section:SNenvelope}

As discussed for instance by \cite{Chevalier05}, rotation-powered pulsars can originate in various types of core-collapse supernov\ae: in Type II supernov\ae{} resulting from red supergiant stars with most of their hydrogen envelope intact (SNIIP), or with most of their hydrogen lost (SNIIL and IIb), or in Type Ib or Type Ic supernov\ae{} (SNIb/c) that stem from stars with all their hydrogen lost. See also \citealp{Maeda07}, \citealp{Woosley10}, and \citealp{Piro11}, \citealp{Kasen10}, for supernov\ae{} associated with magnetars. \cite{Chevalier05} finds that, of the remnants with central rotation-powered pulsars, the pulsar properties do not appear to be related to the supernova category. 

Within a few days after the explosion, the supernova enters a free expansion phase with velocity distribution $v=r/t$, that lasts several hundreds of years. A straightforward way to model the evolution of the density of the ejecta is to assume that the ejected mass $M_{\rm ej}$ will expand spherically in time with a mean velocity $v_{\rm ej}$ over a shell of radius $R_{\rm SN}=v_{\rm ej}t$. The ejected velocity, $E_{\rm ej}$, relates to the supernova explosion energy and the ejected mass through: 
\begin{equation}
v_{\rm ej} = 2\left(\frac{E_{\rm ej}}{M_{\rm ej}}\right)^{1/2}\sim 10^9 \,E_{\rm ej,52}^{1/2}M_{\rm ej,10}^{-1/2}\,{\rm cm\,s}^{-1}\ ,
\end{equation}
where we defined $M_{\rm ej,10}\equiv M_{\rm ej}/10\,{M}_\odot$ and $E_{\rm ej,52}=E_{\rm ej}/10^{52}\,$ergs. Most core-collapse supernov\ae{} are inferred to have explosion energy $E_{\rm ej}\sim 10^{51}\,$ergs.  However, for the pulsars with millisecond to sub-millisecond periods considered here, one can expect that the rotation energy of order $(1/2)I\Omega^2\sim 10^{52}$\,ergs will be transfered within a fraction of a year to the surrounding ejecta (see \citealp{Kasen10}).  Depending on the radiation conversion efficiency of this energy, the surrounding supernova could become ultraluminous. Some ultraluminous SNIb/c and SNII have indeed been detected with an explosion energy $\gtrsim10^{52}\,$ergs (e.g., \citealp{Nomoto01,Woosley10,Piro11,Barkov11}). 

The mean density over $R_{\rm SN}(t)$ can then be written:
\begin{equation}\label{eq:rhoSN}
\rho_{\rm SN}(t)=\frac{M_{\rm ej}}{(4/3)\pi v_{\rm ej}^3t^3} \sim 2\times 10^{-16} M_{\rm ej,10}^{5/2}E_{\rm ej,52}^{-3/2} t_{\rm yr}^{-3}~{\rm g\,cm}^{-3}\,\ ,
\end{equation}
where  $t_{\rm yr}\equiv t/1\,{\rm yr}$, which is the timescale to reach a pulsar spin that enables the acceleration of iron up to $\sim10^{20.5}$~eV (see Eq.~\ref{eq:tspin_simple}). The column density integrated over $R_{\rm SN}$ as a function of time reads
\begin{equation}\label{eq:ySN}
y_{\rm SN}(t) = \rho_{\rm SN}R_{\rm SN} \sim 2\,M_{\rm ej,10}^2E_{\rm ej,52}^{-1} t_{\rm yr}^{-2}~{\rm g\,cm}^{-2}.
\end{equation}

More detailed modelings show that the density evolution of the ejecta is expected to depend on the type of supernova. 
Yet, we demonstrate in what follows that Eq.~(\ref{eq:ySN}) above provides a good estimate for the evolution of the integrated column density of various types of supernova envelopes.  Indeed, we will see in the next section that the escape of UHECRs is determined by their interactions on the baryonic envelopes. Because these interactions solely depend on the integrated column density of the envelope, the detailed density profile is not crucial to our calculations.

Under the assumption of adiabatic, spherically symmetric flows, the numerical calculations of \cite{Matzner99} show that the density of a Type II supernova in the dense central region can take values as high as:
\begin{equation}\label{eq:rhoII}
\rho_{\rm SNII}(t) \sim 10^{-16} M_{\rm ej,10}^{5/2}E_{\rm ej,52}^{-3/2}t_{\rm yr}^{-3}~{\rm g\,cm}^{-3}\,\ .
\end{equation}
Most type II supernov\ae{} eject a mass of order $M_{\rm ej,10}$ \citep{Woosley95}. This dense, relatively flat region extends to radius $R_{\rm b}\sim 2(E_{\rm ej}/M_{\rm ej})^{1/2}t$ and is surrounded by a steep outer power-law profile. The column density that the accelerated particles have to traverse to escape is then:
\begin{equation}\label{eq:yII}
y_{\rm SNII}(t) = \rho_{\rm SNII}R_{\rm b} \sim 4 \,M_{\rm ej,10}^2E_{\rm ej,52}^{-1} t_{\rm yr}^{-2}~{\rm g\,cm}^{-2}.
\end{equation}

For Type Ib/c/bc supernov\ae, one can apply the model of \cite{Matzner99} for the explosion of a star with a radiative envelope, which yields:
\begin{eqnarray}\label{eq:rhoIbc}
\rho_{\rm SNIb/c}(t) &=& 7\times 10^{-17} \left(\frac{v}{0.01c}\right)^{-1.06}\times \nonumber\\
&&M_{\rm ej,2}^{1.97}E_{\rm ej,52}^{-0.97} t_{\rm yr}^{-3}~{\rm g\,cm}^{-3}\ ,
\end{eqnarray}
out to radius $R_{\rm b}$, beyond which the density decreases steeply.
We have assumed in this estimate an explosion energy of $E_{\rm ej,52}$ and an ejecta mass of $M_{\rm ej,2}=M_{\rm ej}/2\,{\rm M}_\odot$, which are derived from the observation of such objects \citep{Drout10}. The corresponding column density, taking into account the velocity distribution $v=r/t$, reads
\begin{equation}\label{eq:yIbc}
y_{\rm SNIb/c}(t) = \int_0^{R_{\rm b}}\rho_{\rm SNIb/c}{\rm d}r \sim 9\,  M_{\rm ej,2}^{2}E_{\rm ej,52}^{-1} t_{\rm yr}^{-2}~{\rm g\,cm}^{-2}\ .
\end{equation}

Equations (\ref{eq:ySN}), (\ref{eq:yII}), and (\ref{eq:yIbc}) agree within factors of a few. It is thus reasonable to consider Eqs.~(\ref{eq:rhoSN}) and (\ref{eq:ySN}) as representative of the envelope mean density and column density, for types II and Ib/c supernov\ae. Equations (\ref{eq:yII}), and (\ref{eq:yIbc}) show that higher ejecta energy $E_{\rm ej}$ and lower masses $M_{\rm ej}$ would enhance the column density. The effects of such cases on particle escape are also discussed throughout the paper.

One can further note that if the pulsar wind shreds its surrounding supernova envelope, as discussed in \cite{Arons03} for the magnetar case, disrupted fragments would expand in the interstellar medium. In this case, one can weight the initial supernova density by $C^{-2/3}$, $C\equiv\delta \rho/\rho$ being a factor measuring the clumpiness of the envelope \citep{Murase09}. A high $C$ would ease the escape of UHECRs from the envelope. However, the values of $C$ remain difficult to evaluate, as no observational evidence of such phenomena has been detected.\\

The composition of the supernova ejecta depends upon the type, progenitor
mass, and the final interior mass of the supernova.  CXO J164710.2-455216's
association with the Westerlund 1 star cluster argues that at least some
pulsars arise from massive star progenitors \citep{ Muno06}.  But, as mentioned before, rotation-powered pulsars and magnetars have been invoked for a wide variety of supernova types. The composition of a type Ib supernova is roughly 50\% helium and
50\% C/O:  e.g., the \cite{Woosley10} progenitor is $\sim$50\% helium,
$\sim$43\% carbon and $\sim$7\% oxygen.  Type Ic supernovae (more numerous
than Ib supernovae) are composed almost entirely of C/O and heavier
elements:  e.g., \cite{Mazzali10} argued that SN 2007gr was composed
of roughly 75\% C, 15\% O, 8\% $^56$Ni, and 2\% S.  Type II supernovae have a
range of ejecta, ranging from roughly 60\% H, 30\% He, and 10\% C/O to
explosions very similar to type Ib supernova with small amounts of H.

We will discuss in Section 3.4 how the escaped UHECR spectrum varies
between pure hydrogen and pure helium envelopes (or helium
and carbon envelopes).

\subsection{Analytical estimates}\label{subsection:analyt_escape}

\begin{figure*}
\centering
\epsfig{file=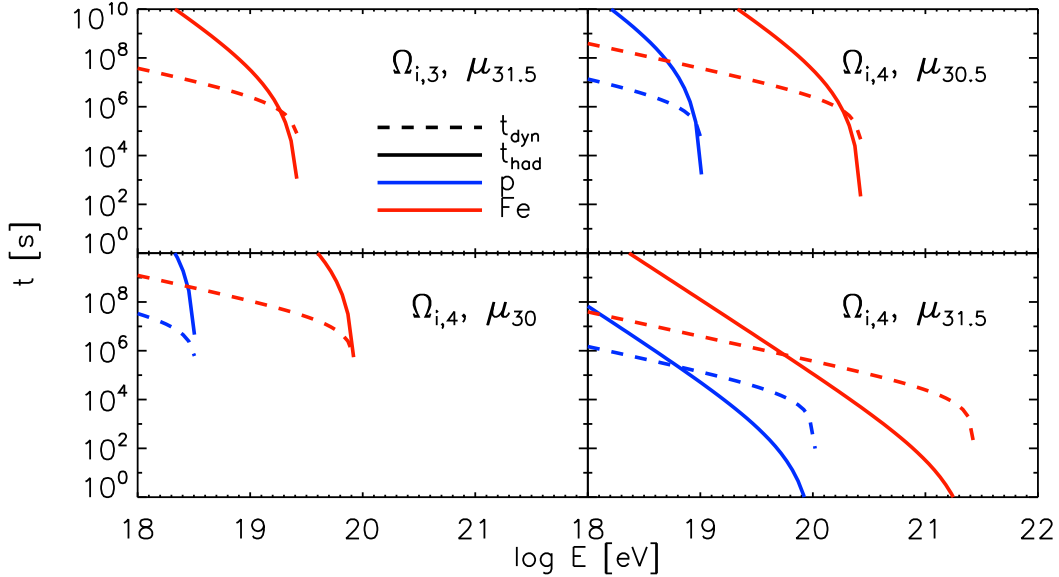,width=0.8\textwidth}
\caption{Timescales at play for the escape of UHECRs from a supernova envelope with $M_{\rm ej,10}$ and $E_{\rm ej,52}$. The crossing time $t_{\rm dyn}$ (dashed lines) and energy loss time by hadronic interactions $t_{\rm had}$ (solid lines) are displayed as a function of particle energy $E$, for pure iron (red) and pure proton (blue) injections. The timescales are calculated for various pulsar initial rotation velocities $\Omega_{\rm i}=10^{3},10^4\,$s$^{-1}$ and magnetic dipole moments $\mu=10^{30}$ to $10^{31.5}\,$cgs, as labeled. The other pulsar parameters are set to $I=10^{45}\,$g\,cm$^2$, $\eta=0.1$.  }
\label{fig:tloss}
\end{figure*}

In accord with the discussion at the beginning of Section~\ref{section:SNenvelope}, we will consider in the following that Eqs.~(\ref{eq:rhoSN}) and (\ref{eq:ySN}) provide a reasonable estimate of the evolution of the density of the supernova envelope surrounding the neutron star.

Successful escape of UHECRs from the envelope will occur if the shell crossing time $t_{\rm dyn}$ is shorter than the cooling time by hadronic, $t_{\rm had}$, and photo-hadronic, $t_{N\gamma}$, interactions.

The acceleration of a particle to the energy $E$ happens at a time after pulsar birth:
$t(E)\simeq t_{\rm spin}(E)$.
We can assume that the thickness of the supernova shell to traverse at a time $t$ is given by $R_{\rm SN}\simeq tv_{\rm ej}$. Indeed, from the values given in Section~\ref{subsubsection:site_beyond}, the acceleration site $R_{\rm a}\ll R_{\rm SN}$, as soon as $t\gtrsim100$~s. The crossing time for UHECRs traveling at the speed of light then reads:
\begin{equation}\label{eq:tdyn}
t_{\rm dyn}(E) \simeq\frac{R_{\rm SN}}{c}\simeq \frac{v_{\rm ej}}{c}t_{\rm spin}(E) \ .
\end{equation}
As the expansion time scale of the envelope is $t_{\rm ex}=R_{\rm SN}/v_{\rm ej}\simeq t_{\rm spin} < t_{\rm dyn}$, one can neglect the evolution of the envelope density during the escape of a particle.

The timescale for hadronic interaction losses can be expressed as:
\begin{equation}\label{eq:thadloss}
t_{\rm had}(E) = m_{\rm b}\{c \,\rho_{\rm SN}[t_{\rm spin}(E)]\sigma(E)\xi(E)\}^{-1} \ ,
\end{equation}
where $m_{\rm b}$ is the mass of the dominant target ion composing the envelope. The parameters $\xi(E)$ and $\sigma(E)$ are the elasticity and the cross-section of the interaction at energy $E$. For our analytical estimates, we evaluate their values roughly from the hadronic interaction model EPOS \citep{WLP06}. We assume that the cross-sections of the hadronic interactions do not vary strongly above $E>10^{18}~$eV and set them to $\sigma_p=130\,$mb for proton-proton interactions and $\sigma_{\rm Fe}=1.25\,$b for iron-proton interactions. The number of nucleons carried out at each interaction can vary from 1 to $A-1$, with a large spread in values. For demonstrative purposes, we take an average value of $\xi=0.4$ for both $p-p$ and $p-$Fe interactions. These calculations are done accurately using EPOS in our numerical calculations in the next section.

The condition of escape from the supernova envelope can thus be written as $t_{\rm dyn}/t_{\rm had}<1$, yielding 
\begin{eqnarray}
t&>&t_{{\rm esc},p}\sim 1.2\times 10^7\, M_{\rm ej,10}E_{\rm ej,52}^{-1/2}\,{\rm s}  \quad \mbox{for proton},\label{eq:tcrossthad1}\\
t&>&t_{\rm esc,Fe} \sim3.8\times 10^7\, M_{\rm ej,10}E_{\rm ej,52}^{-1/2} \, {\rm s} \quad \mbox{for iron},\label{eq:tcrossthad2}
\end{eqnarray}
where we assumed a supernova density profile following Eq.~(\ref{eq:rhoSN}).
Cosmic rays at ultrahigh energy will escape only if they are produced at late times $t\gtrsim1$~yr, when the envelope density has decreased.
Because nuclei of charge $Z$ at a given energy $E$ are produced at a time $t_{\rm spin}(E)\propto Z$ (Eq.~\ref{eq:tspin_simple}), one has $t_{\rm dyn}/t_{\rm had}\propto Z^{-2}$.
The escape condition from the baryonic envelopes at a fixed $E$ should consequently be eased for heavier nuclei.

Still assuming the supernova density profile of Eq.~(\ref{eq:rhoSN}), and using the spin-down time given in Eq.~(\ref{eq:tspin_simple}), one can express the cut-off energy above which injected primary particles should not be able to escape the envelope:
\begin{eqnarray}\label{eq:Ecut_estimate}
&&E_{{\rm cut},Z} =E_{\rm i}\,\left[1+{8\over9}{{\mu^2\Omega_{\rm i}^2}\over{Ic^3}}\left( \frac{3M_{\rm ej}\sigma\xi}{4\pi m_{\rm b}v_{\rm ej}^{2}} \right)^{1/2}\right]^{-1}\\
&&\sim  7.5\times10^{18}\,Z_{1}\eta_1I_{45}\mu_{30.5}^{-1}M_{\rm ej,10}^{-1}E_{\rm ej,52}^{1/2}\left(\frac{\sigma_p}{\sigma}\right)^{1/2}\,{\rm eV} \label{eq:Ecut_proton} \\
&&\sim 1.2\times 10^{20}Z_{26}\eta_1I_{45}\mu_{30.5}^{-1}M_{\rm ej,10}^{-1}E_{\rm ej,52}^{1/2}\left(\frac{\sigma_{\rm Fe}}{\sigma}\right)^{1/2}{\rm eV} \label{eq:Ecut_iron}
\end{eqnarray}
where the first numerical application corresponds to protons and the second to iron nuclei. Note that under the crude approximation that $\sigma\propto A^{2/3}$, $E_{\rm cut,Z}\propto Z/A^{1/3}$. For $E_{{\rm cut},Z} \ll E_{\rm i}$, $E_{\rm cut,Z}$ does not depend on $\Omega_{\rm i}$. 

This trend is illustrated in Figure~\ref{fig:tloss}, where the main timescales at play are displayed: $t_{\rm dyn}$ and $t_{\rm had}$ as a function of particle energy $E$, for various pulsar parameters $\Omega$ and $\mu$, and for both pure iron and pure proton injections.
As expected, iron particles can escape the envelope at higher energies, as they can reach these energies at later times.
Lower magnetic fields ($\mu\lesssim 10^{31}$) lead to longer $t_{\rm spin}$ at a fixed $E$ (Eq.~\ref{eq:tspin_simple}), while  high pulsations ($\Omega\gtrsim 10^4\,$s) lead to higher acceleration energies (Eq.~\ref{eq:Eacc}). 

When iron nuclei are injected, secondary particles are produced by hadronic interactions for times $t<t_{\rm esc,Fe}$. These secondaries of mass and charge numbers ($A,Z$) can escape the envelope only at times $t>t_{{\rm esc,}Z}$, where $t_{{\rm esc,}Z}$ is defined as the time at which $t_{\rm dyn}/t_{\rm had}=1$. Hence, secondaries that will escape from the envelope have necessarily been produced between $t_{\rm esc,Z}<t<t_{\rm esc,Fe}$, i.e., the lightest secondaries will escape first. This translates in terms of the energy range of the primary iron to: $E_{\rm cut,Fe}<E<E_{\rm Fe}(t_{{\rm esc,}Z})$, where we can further express $E_{\rm Fe}(t_{{\rm esc,}Z})=(26/Z)E_{{\rm cut,}Z}$. The main fragment among secondary particles will thus emerge from the envelope between energies $E'_{{\rm low},Z}\lesssim E\lesssim E'_{{\rm cut,}Z}$, with 
\begin{eqnarray}
E'_{{\rm low},Z} &\equiv& \frac{A}{56}E_{\rm cut,Fe} \\ \nonumber
 &\sim& 2.1\times10^{18} A \eta_1I_{45}\mu_{30.5}^{-1}M_{\rm ej,10}^{-1}E_{\rm ej,52}^{1/2}\,{\rm eV}\ ,\label{eq:sec_cutoff} \\
E'_{{\rm cut},Z} &\equiv& \frac{26}{56}\frac{A}{Z}E_{{\rm cut,}Z}\\  \nonumber
&\sim& 3.5\times 10^{18} A \eta_1I_{45}\mu_{30.5}^{-1}M_{\rm ej,10}^{-1}E_{\rm ej,52}^{1/2}\,{\rm eV}.
\end{eqnarray}

The numerical estimates are calculated for secondary protons. Peaks of the various secondary elements should appear in the escaped cosmic-ray spectrum at their respective energies.
A tail due to lower energy secondary nucleons ($E<E'_{{\rm low},Z}$) following approximately a power-law with index $\sim -1/2$ should also be produced together with the main fragment, down to PeV energies. The amplitude of this tail around $\sim E'_{{\rm low},Z}$ is about a fraction of the number of the main fragment.

From Eq.~(\ref{eq:tdyn}) and (\ref{eq:thadloss}), one can derive: $t_{\rm dyn}/t_{\rm had}\sim 3\times 10^{10}\, t_{2}^{-2} M_{\rm ej,10}^{2}E_{\rm ej,52}^{-1}$ at $t_2\equiv t/100\,$s, for $(A,Z)=(90,40)$. We assumed a cross-section $\sigma_{\rm 90}=1.5\,$b for nuclei-proton interactions and an elasticity of $\xi=0.4$, at energies $E\sim10^{20}$~eV (in the target rest-mass frame).
This demonstrates that nuclei with $A\gtrsim 56$ that could be injected at times $t\sim 10-100$~s if a successful $r$-process occurred in the neutrino-driven wind (Section~\ref{subsection:injection}), cannot survive the crossing of the supernova envelope.\\

Ultrahigh energy ions could also experience photo-disintegration in the radiation fields generated at the interface between the pulsar wind and the supernova shell. 

This radiation field can be expected to be significant if the supernova explosion is driven by the pulsar wind, as expected for millisecond rotators. A fraction $\eta_\gamma$ of the wind energy $\sim (1/2)I\Omega_{\rm i}^2$ can be converted to radiative energy via internal shocks and another fraction $\eta_{\rm th}$ of this radiation then thermalizes depending on the opacity of the medium. This thermal component peaks at energy $\epsilon_\gamma = kT \sim 0.4\,(\eta_{\gamma,1}\eta_{\rm th})^{1/4} E_{\rm ej,52}^{-1/8}M_{\rm ej,10}^{3/8}t_{\rm yr}^{-3/4}\,$eV, with energy density $U_{\rm th}\sim 0.5\,\eta_{\gamma,1}\eta_{\rm th}E_{\rm ej,52}^{-1/2}M_{\rm ej}^{3/2}t_{\rm yr}^{-3}\,$erg\,cm$^{-3}$, where $\eta_{\gamma,1}\equiv \eta_\gamma/0.1$. This background leads to a cooling time by photo-disintegration of order:
\begin{eqnarray}
t_{A\gamma,{\rm th}} &=&  [c\,\xi_{A\gamma}(\Delta\epsilon_{A\gamma}/\bar{\epsilon}_{A\gamma})\sigma_{A\gamma}U_{\rm th}/\epsilon_\gamma]^{-1}\\
&\sim& 10^5 A_{56}^{-0.21} \left(\frac{E_{\rm ej,52}}{\eta_{\gamma,1}^{2}\eta_{\rm th}^{2}}\right)^{3/8} M_{\rm ej,10}^{-9/8}t_{\rm yr}^{9/4} \,{\rm s}
\end{eqnarray}
where $\Delta\epsilon_{A\gamma}/\bar{\epsilon}_{A\gamma}\sim 0.4\,A_{56}^{0.21}$, $\sigma_{A\gamma}\sim 8\times 10^{-26}\,A_{56}\,$cm$^{-2}$ \citep{Murase08}, and we take for the elasticity of the $A\gamma$ interaction: $\xi_{A\gamma}=1/A$ (which is a crude approximation). This estimate of the cooling time is valid for cosmic-ray energy $E_{A,{\rm peak}}\sim 4\times10^{17}\,(\eta_{\gamma,1}\eta_{\rm th})^{-1/4} E_{\rm ej,52}^{1/8}M_{\rm ej,10}^{-3/8}t_{\rm yr}^{3/4}\,$eV, and is about one order of magnitude larger for $E_A\gtrsim E_{A,{\rm peak}}$, as the photo-disintegration cross-section lowers. At the highest energies ($E_A\sim 10^{20}$\,eV), photo-disintegration could thus play a role on the escape of cosmic rays if the radiation and thermalization efficiencies are higher than $\eta_\gamma\eta_{\rm th}\gtrsim10^{-2}$. The rate of wind energy going to radiation is evaluated to be of order 10\% (e.g., \citealp{Kasen10}), but the thermalization fraction of these photons, $\eta_{\rm th}$, is not known, due to the uncertainties on the opacities in the internal shock region. Mixing and Rayleigh-Taylor instabilities effects creating finger-type structures could lead to a leaking of the high energy photons, and the thermalization fraction could be as low as $\lesssim10\%$. A higher acceleration efficiency $\eta$ would also enable particles to reach the highest energies by the time the radiation field intensity has become negligible. Given these uncertainties, and for simplicity, we will assume in this paper that the radiation field can be neglected for the escape of UHECRs from supernova envelopes, the baryonic background playing the major role. \\

To summarize, the conditions for successful acceleration and escape above $10^{20}\,$eV can be written as:
\begin{eqnarray}\label{eq:param_scan}
\left\{
\begin{array}{ll}
B\Omega_{\rm i}^2 \gtrsim  (10^{12.4}\,{\rm G})\times (10^4\,{\rm s}^{-1})^2 \,Z_{26}^{-1}\eta_1^{-1}R_{*,10}^{-3}\\
B\lesssim10^{12.8}\,{\rm G}\,Z_{26}A_{56}^{-1/3}\eta_1I_{45}M_{\rm ej,10}^{-1}E_{\rm ej,52}^{1/2}R_{*,10}^{-3}
\end{array}
\right.
\end{eqnarray}
Higher values of the magnetic field would allow higher acceleration energies, but would require lower ejecta mass and higher explosion energies. Note that $10\,M_\odot$ can be viewed as an upper bound for the ejecta mass for type II supernov\ae{} \citep{Woosley02}. One might also advocate that the presence of clumps could lower the overall densities and allow the escape of particles at $E>10^{20}$~eV. All in all, the parameter space allowed for successful acceleration and escape appears to be narrow, but we will see in Section~\ref{subsection:flux} that the low rate of sources required to account for the observed UHECR flux would compensate for this issue. A higher acceleration efficiency $\eta$ would also broaden the allowed parameter space.

\subsection{Numerical Setup}\label{subsection:setup}

As discussed in the previous section, the hadronic interaction between UHECRs and the baryonic envelopes is the determinant factor that would affect the injected UHECR spectrum.

The interactions with the baryonic envelopes were calculated by Monte-Carlo for injected nuclei and their secondaries.
 As in \cite{Kotera09}, we used the hadronic interaction model
EPOS \citep{Werner06} and the fragmentation model of \cite{Campi81}, as implemented in the air shower
simulation code CONEX \citep{Bergmann07}.

In the case of  a non-hydrogen baryonic envelope, the interaction products can be derived from the nuclei-proton interaction case by a superposition law. In the target rest frame, the products of the interaction between a projectile of mass number and energy $ (A_{\rm proj},E_{\rm proj})$ and a target nucleus of mass number $A_{\rm targ}$ are roughly equivalent to $A_{\rm targ}$ times the products of the interaction between a projectile with $(A_{\rm proj},E_{\rm proj}/ A_{\rm targ})$ and a target proton. The exact cross-sections are nonetheless computed with EPOS.

In the simulations, we modeled pulsars with initial angular velocity $\Omega_{\rm i}\sim 10^{3.0-4.2} \,$s$^{-1}$ and magnetic moment $\mu\sim 10^{30-33} \, \rm cgs$, corresponding to a surface magnetic dipole field $B\sim 2\times  10^{12-15}\,\rm G$. Notice that there is an upper limit ($\sim10^{4.2}\,\rm s^{-1}$) on the initial angular speed \citep{Haensel99}. 
For each set of parameters, $10^7$ cosmic rays  are injected following a power-law energy spectrum as in Eq.~(\ref{eq:spectrum_arons}) with minimum injection energy $E_{\rm min}=10^{17}\,\rm eV$, and the maximum acceleration energy $E_{\rm i}$ calculated in Eq.~(\ref{eq:Eacc}). Above $E_{\rm i}$, the spectrum cuts-off exponentially. Nuclei with initial energy $E$ are injected at a radius $R_{\rm a}=10^{10}\;\rm cm$ (corresponding to $\sim 3\,r_{\rm min}$ for $\Omega_{\rm i}=10^4\,$s$^{-1}$, see discussion in Section~\ref{subsubsection:site_beyond}) at the time $t_{\rm spin}(E)$, and propagate through a supernova envelope of total ejected mass  $10\,M_{\odot}$ ($2\,M_{\odot}$ in Type Ib/c supernova case) expanding at a constant rate $v_{\rm ej}=10^9 \,E_{\rm ej,52}^{1/2}M_{\rm ej,10}^{-1/2}\,{\rm cm\,s}^{-1}$. The evolution of the ejecta density is assumed to follow Eq.~(\ref{eq:rhoSN}).
We studied pulsars embedded in pure hydrogen, helium and carbon supernov\ae.

\subsection{Numerical Results}\label{subsection:numerical}

We first assume a pure hydrogen envelope. The results are presented in Section~\ref{subsubsection:hydrogenenvelope}. Simulations using more supernova envelopes with heavier composition are discussed in Section~\ref{subsubsection:realenvelope}.

\subsubsection{Pure Hydrogen supernova envelope}\label{subsubsection:hydrogenenvelope}

\begin{figure}
\centering
\epsfig{file=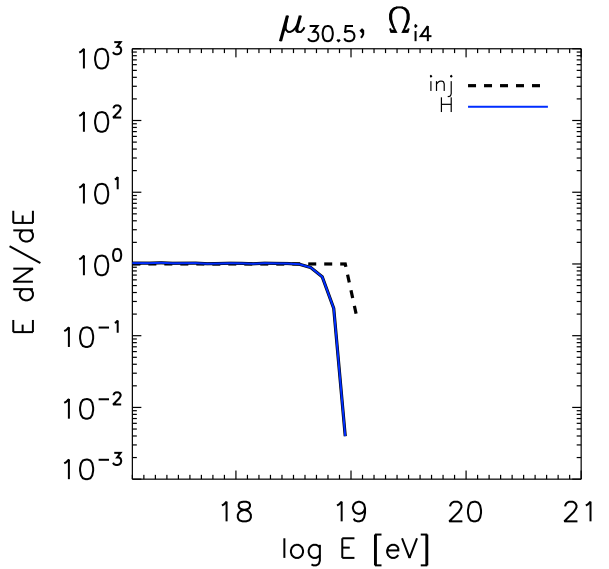,width=0.5\textwidth}
\caption{\label{fig:proton} UHECR spectrum before (dash) and after (solid) escape from a hydrogen supernova envelope with $M_{\rm ej,10}$ and $E_{\rm ej,52}$, with pure proton injection. The pulsar parameters are $I=10^{45}\,$g\,cm$^2$, $\eta=0.1$, $\Omega=10^{4.0}\,\rm s^{-1}$, and $\mu= 10^{30.5}\, \rm cgs$.}
\end{figure}

\begin{figure}
\centering
\epsfig{file=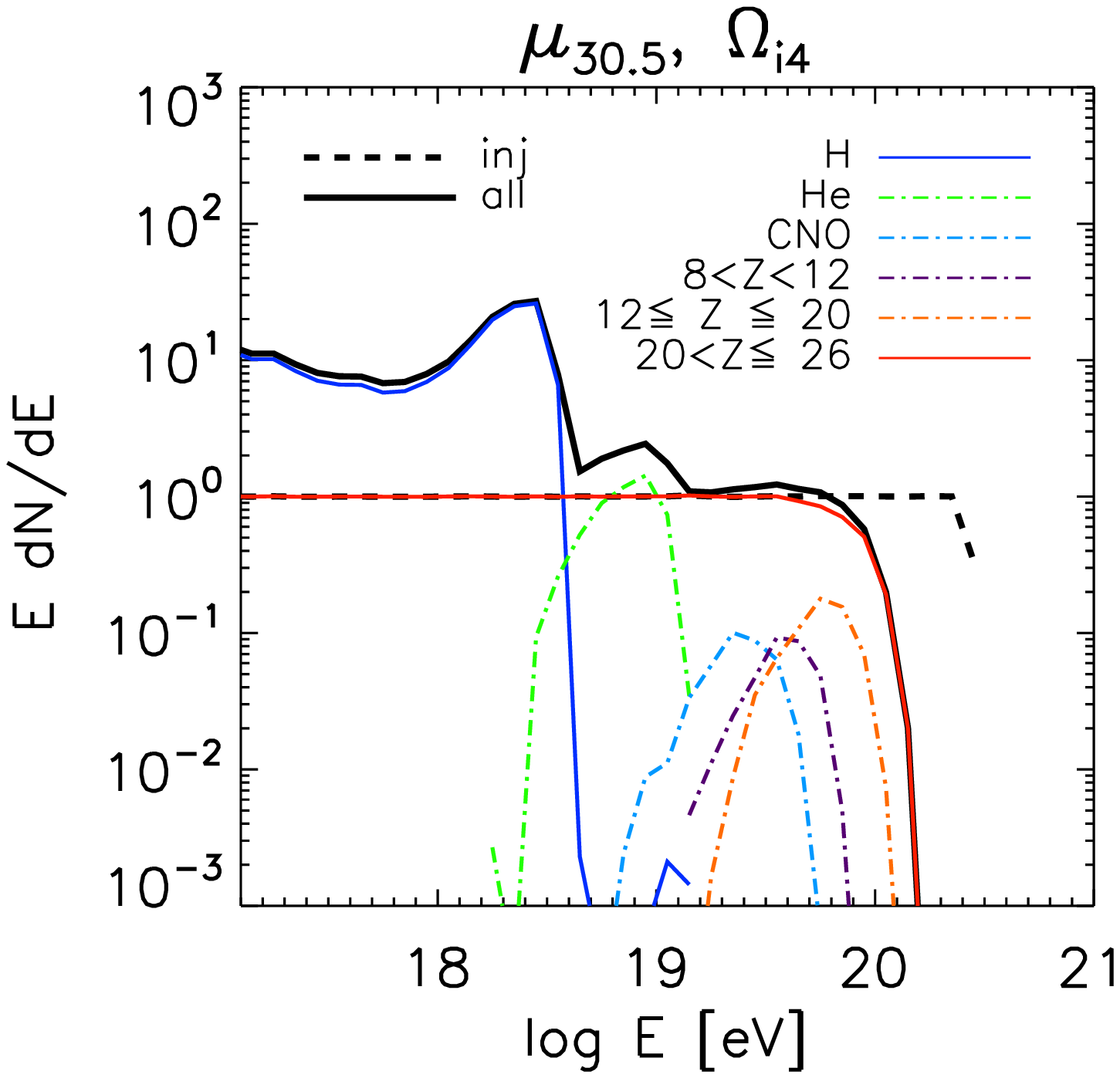,width=0.5 \textwidth}
\epsfig{file=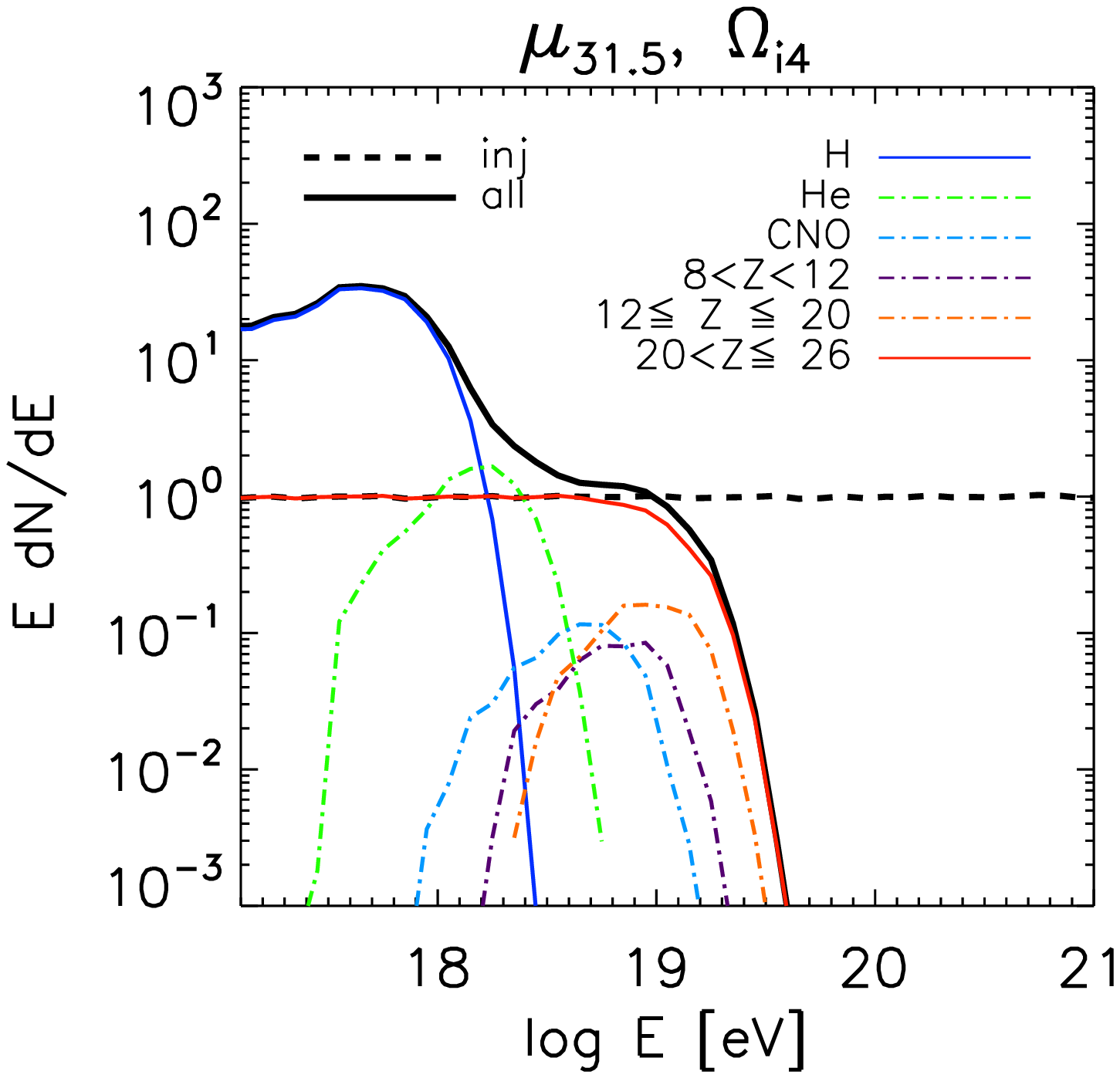,width=0.5 \textwidth}
\caption{\label{fig:iron} UHECR spectrum before (dash) and after (solid and dash dotted) escape from hydrogen supernova envelope with $M_{\rm ej,10}$ and $E_{\rm ej,52}$, with pure iron injection. The pulsar parameters are $I=10^{45}\,$g\,cm$^2$, $\eta=0.1$, $\Omega=10^{4}\,\rm s^{-1}$, and $\mu= 10^{30.5}\,$cgs (top), and $\mu=10^{31.5}\,\rm cgs$ (bottom). Different compositions are listed as in the legend box. }
\end{figure}

Figure~\ref{fig:proton} presents the injected (in dash line) and escaped  (in solid line) spectra of pure proton injection by a pulsar with initial angular speed $\Omega_{\rm i}=10^{4}\,\rm s^{-1}$ and magnetic dipole moment $\mu=10^{30.5}\,\rm cgs$. The injected spectrum follows the characteristic $ -1$ spectral index in Eq.~(\ref{eq:spectrum_arons}).  As predicted in Eq.~(\ref{eq:Ecut_proton}) UHE protons above $\sim 10\,\rm EeV$ fail to escape the supernova envelope, since the region is still very dense at the time they are produced. Below a few EeVs protons are free to escape. Protons with energy in between can partially escape with significant flux suppression. EPOS shows that for one $10\,\rm EeV$ primary proton, the peak of interaction products lies at $10^{14}\,\rm eV$; the chance of resulting a secondary proton with  $E\ge 10^{17}\,\rm eV$   is less than $0.01$. Therefore we can barely see the secondary protons in our energy window of simulation.

The spectra of pure iron injection by pulsars with $\Omega_{\rm i}=10^{4}\,\rm s^{-1}$ and $\mu=10^{30.5},10^{31.5}\,\rm cgs$ are shown in Fig~\ref{fig:iron}. In the top plot ($\mu=10^{30.5}\,\rm cgs$, $\Omega=10^4\,\rm s^{-1}$), primary iron nuclei with energy up to $E_{\rm cut, Fe}=1.2\times10^{20}\,\rm eV$ can escape without significant loss. As discussed in our analytical estimates,  most secondaries should originate from primary iron nuclei with energy between $E_{\rm cut, Fe}=1.2\times 10^{20}\,\rm eV$ and $56\times E_{\rm cut, p}=4.2\times10^{20}\,\rm eV$, corresponding to the iron cutoff and iron mass number times  the cutoff of  secondary protons. In agreement with Eq.~(\ref{eq:sec_cutoff}), secondaries lie between $(1.0-5.0)\times10^{18}\,\rm eV$ for proton, $2.0\times 10^{18} - 1.3\times10^{19}\,\rm eV$ for helium, $7.9\times10^{18} - 4.0\times10^{19}\,\rm eV$ for CNO,  $(1.3-7.1)\times 10^{19}\,\rm eV$ for Mg-like elements and $2.0\times 10^{19} - 1.1\times 10^{20}\,\rm eV$ for Si-like elements, with the peak positions scaled to the mass number of the elements and the bump width being almost the same in logarithmic coordinates. The significant tail of protons below $1\,\rm EeV$ comes from the products of the hadronic interactions. On average, each interaction of a $500\,\rm EeV$ iron nucleus results in one EeV proton among its  products. The strong signals from secondary nuclei contribute to a steeper overall spectrum (in solid black) which follows $\sim E^{-2}$ at $10^{18.5}-10^{20}\,\rm eV$.

When the magnetic field is $10$ times stronger ($\mu=10^{31.5}\,\rm cgs, \Omega_{\rm i}=10^4\,\rm s^{-1}$, bottom plot of Fig~\ref{fig:iron}), the pulsar spins faster and the cutoff for primary and secondaries are lowered by $10$ times (see Eq.~\ref{eq:sec_cutoff}). Hence, the $\mu=10^{31.5}\,\rm cgs$ case presents a similar shape as the $\mu=10^{30.5}\,\rm cgs$ case except an overall shift to lower energies by a factor of 10.

\begin{figure}
\centering
\epsfig{file=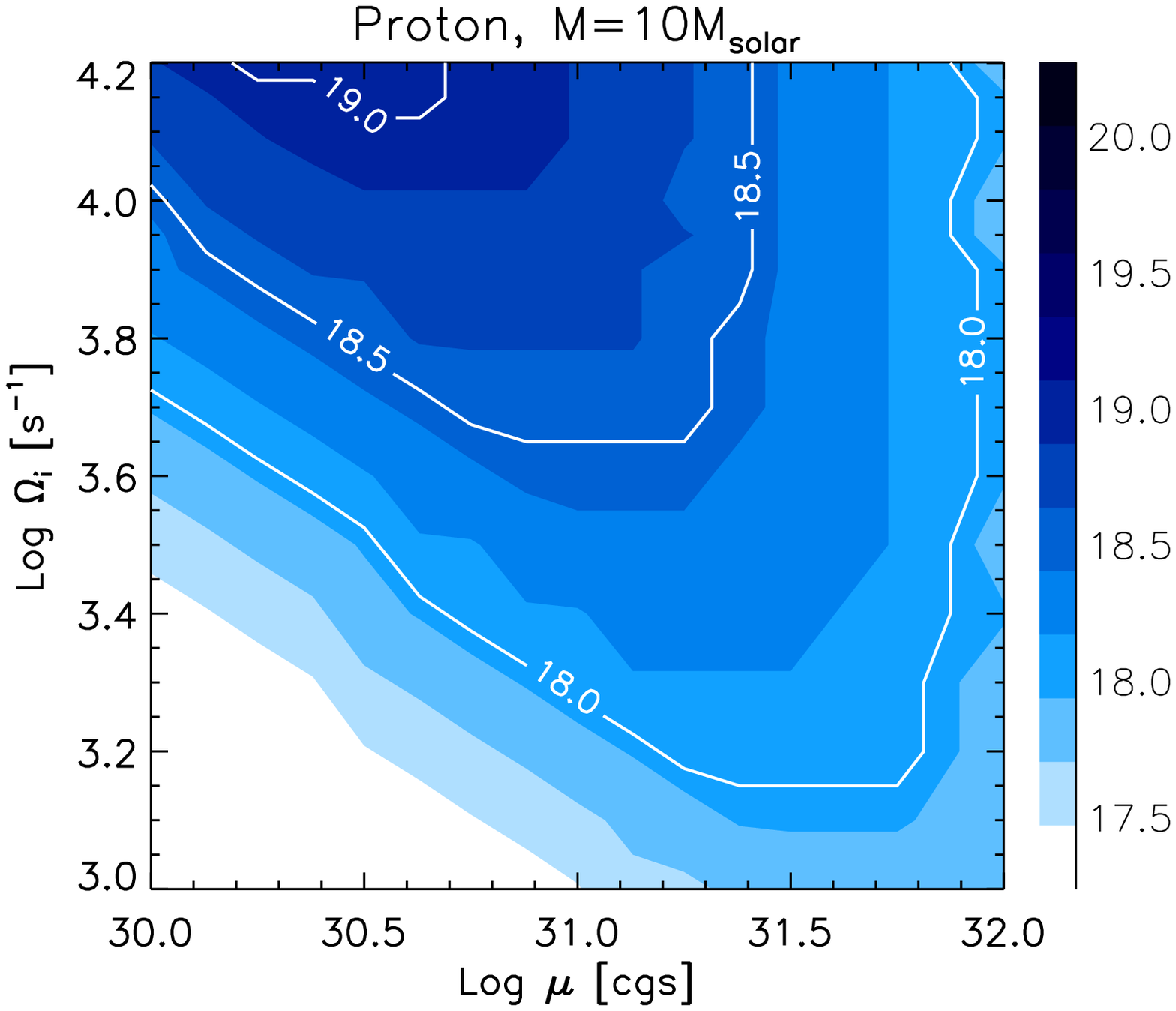,width=0.5\textwidth}
\epsfig{file=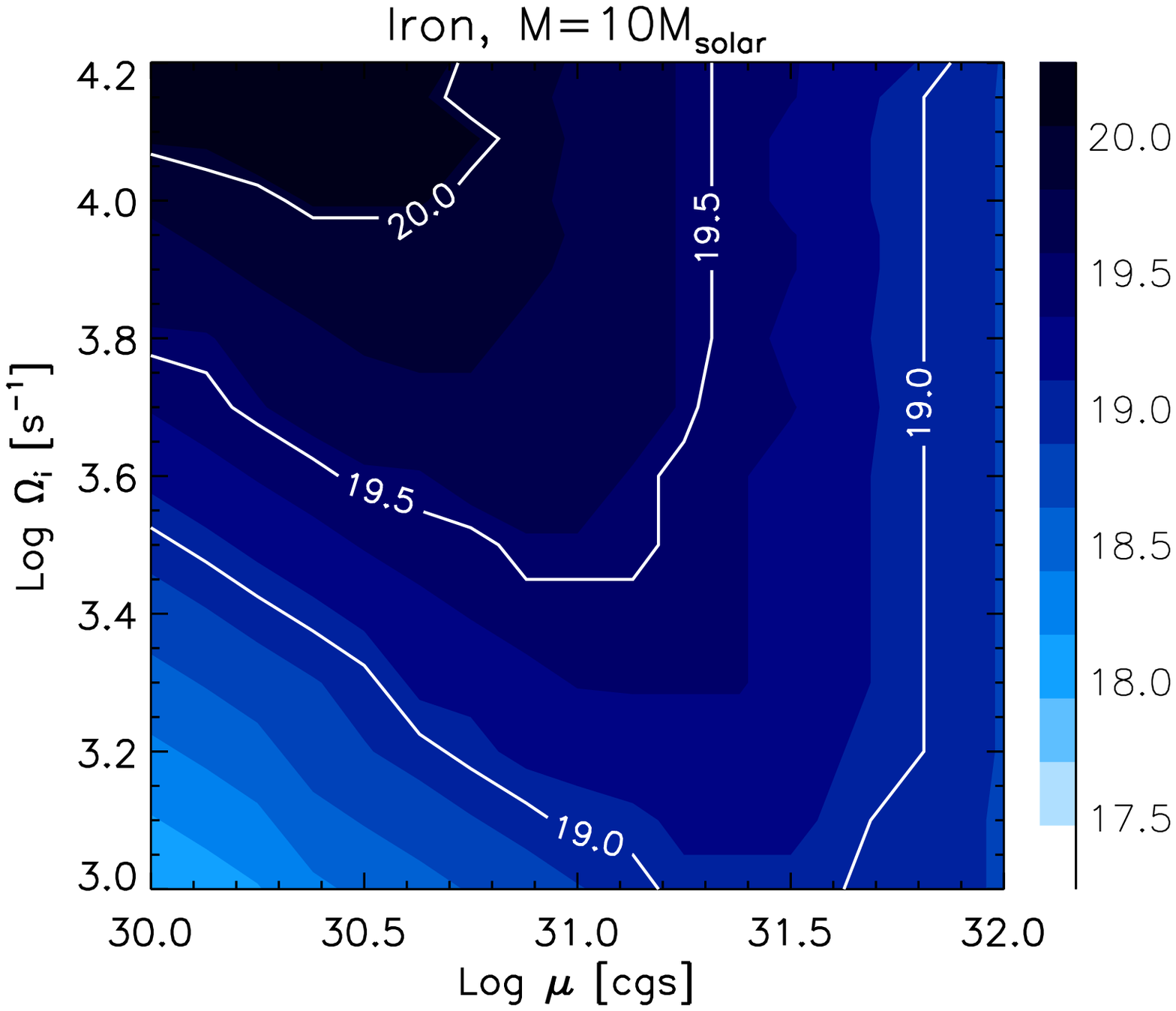,width=0.5\textwidth}
\caption{\label{fig:contours} Parameter space with cut-off energy ($E_{\rm cut,Z}$) contours, for a hydrogen supernova envelope with $M_{\rm ej,10}$ and $E_{\rm ej,52}$, and pulsar parameters $I_{45}$ and $\eta_1$. The solid lines refer to cut-off particle energies after the escape. Up is proton injection and down is iron injection. Notice that current neutron star models suggest an upper limit of rotational speed at $\Omega_{\rm i}\le 10^{4.2}\,\rm s^{-1}$. Note also that $E_{\rm cut,Z}$ scales with $M_{\rm ej,10}^{-1}E_{\rm ej,52}^{1/2}$ (Eq.~\ref{eq:Ecut_estimate}).}
\end{figure}

As pointed out in  Section~\ref{subsection:unipolar}, at low energies when $E \ll E_{\rm g}$ the gravitational wave losses are negligible and $t_{\rm spin}$ is independent on the initial rotation speed $\Omega_{\rm i}$ for $E \ll E_{\rm i}$.  
A pulsar with higher initial angular velocity can inject UHECRs with greater maximum energy.   However a minimum spin period $\sim 0.4\;\rm ms$ is allowed for neutron stars \citep{Haensel99} corresponding to an upper limit ($\sim10^{4.2}\,\rm s^{-1}$) on the initial angular speed. Magnetic dipole moments $\mu$ greater than $10^{32} \,\rm cgs$ would make the spin-down process too fast to allow UHECR escape. On the other hand pulsars with $\mu<10^{30}\,\rm cgs$ are not energetic enough to accelerate particles to ultrahigh energy (see Eq.~\ref{eq:param_scan}). To determine the best escaping region we ran a parameter scan with $15\times15$ sets of $(\Omega,\mu)$ and the results are presented in Figure~\ref{fig:contours}.

We define the cut-off energy $E_{\rm cut}$ as the energy the ratio between the escaped and injected particles is less than $10\%$. It  corresponds approximately to the highest energy of escaped cosmic rays $E_{\rm cut, Z}$ defined in Eq.~(\ref{eq:Ecut_proton}).  In Figure~\ref{fig:contours}, the contours represent $E_{\rm cut}$ reached after escaping hydrogen supernova envelopes with $M_{\rm ej,10}$ and $E_{\rm ej,52}$ for pulsars with dipole moment $\mu$ and initial angular velocity $\Omega_{\rm i}$. In the proton case (top), protons with energy above $10^{20}\,\rm eV$ cannot escape the supernova envelope in our model. In the iron contours (bottom), the parameter region with $(\mu\approx10^{30.00-30.72}\,\rm cgs)\times (\Omega_{\rm i}\approx 10^{3.95-4.20}\,\rm s^{-1})$ allows the escape of  iron nuclei with energy greater than $10^{20}\,\rm eV$. This parameter scan is based on a supernova envelope with density profile described in Eq.~({\ref{eq:rhoSN}}). Higher values of explosion energy and lower ejecta mass could lead to a broader enclosed parameter region that allows the escape, as $E_{\rm cut,Z}$ scales with $M_{\rm ej,10}^{-1}E_{\rm ej,52}^{1/2}$ (Eq.~\ref{eq:Ecut_estimate}). Our results agree with the theoretical prediction from Fig.1 in \cite{Blasi00}, except that we have a smaller parameter area that allows escape. This comes from our assumption that only $\eta\sim 10\%$ of the induced potential turns into UHECR energy.

\subsubsection{Helium-Carbon/Hydrogen-Helium supernova envelopes}\label{subsubsection:realenvelope}

\begin{figure}[t]
\centering
\includegraphics[width=0.5\textwidth]{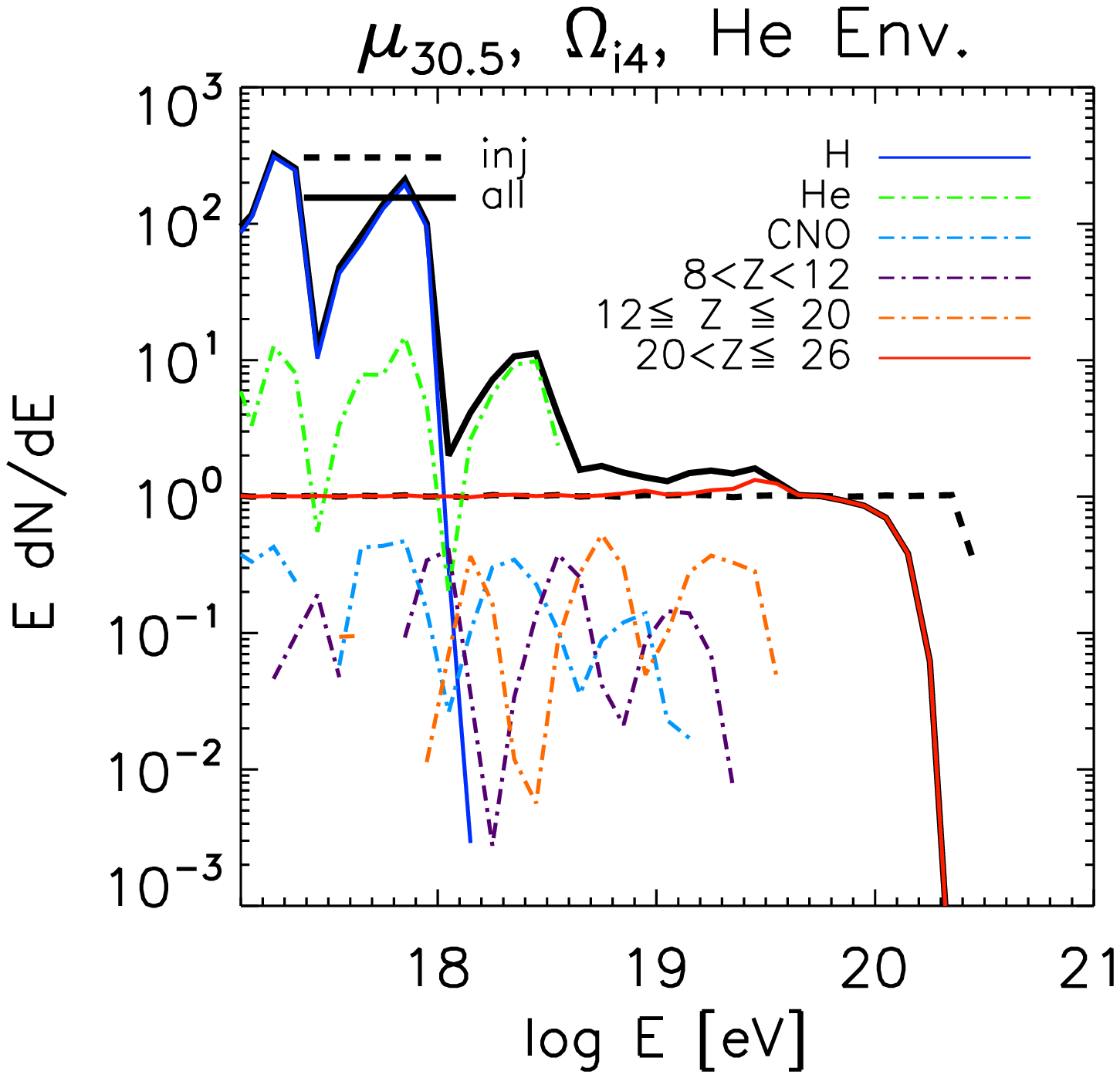}
\includegraphics[width= 0.5\textwidth]{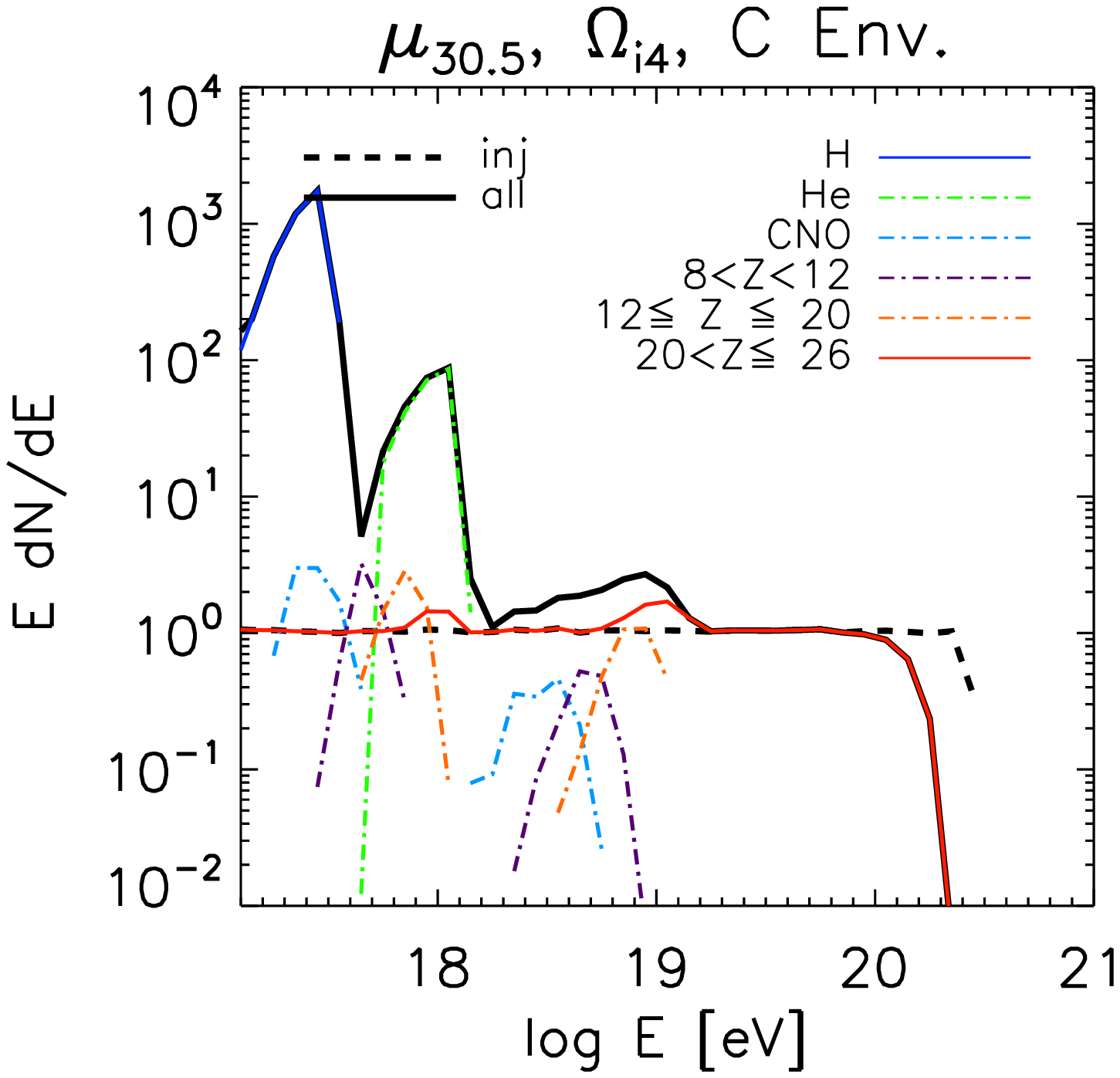}
\caption{\label{fig:real} UHECR spectrum after escape from a supernova envelope with $M_{\rm ej,10}$ and $E_{\rm ej,52}$, with composition ({\it top}): $100\%$ $^{4}\rm He$ and ({\it bottom}): $100\%\;^{12} \rm C$. The pulsar parameters are $I=10^{45}\,$g\,cm$^2$, $\eta=0.1$, $\Omega=10^{4}\,\rm s^{-1}$, and $\mu= 10^{30.5}\,$cgs.}
\end{figure}

Results in Figure~\ref{fig:real} are from simulations with an ejecta mass of $M_{\rm ej}=10\,M_\odot$, explosion energy $E_{\rm ej}=10^{52}\,$ergs, and a composition of pure $^4\rm He$ (top) and pure  $^{12}\rm C$  (bottom). As discussed in Sec.~\ref{subsection:setup}, realistic envelopes for SNII and SNIb/c are more complicated and could be evaluated by a combination of Fig.~\ref{fig:iron} and \ref{fig:real}.  Spectra of UHECRs escaped from envelopes abundant in heavier elements maintain features from that with a pure hydrogen  envelope. For instance, in case of a pure helium envelope (top plot in Fig.~\ref{fig:real}), the spectrum preserves the `original' secondary peaks at $6.3\times10^{17}\,\rm eV$ for hydrogen, $2.5\times10^{18}\,\rm eV$ for helium, $8.0\times10^{18}\,\rm eV$  for CNO,  $1.3\times 10^{19}\,\rm eV$ for Mg-like elements and $2.5 \times 10^{19}\,\rm eV$ for Si-like elements. These peaks are similar to the ones in a pure hydrogen envelope (see the first plot of Fig.~\ref{fig:iron}), except that they are located at $4$ times lower in energy, due to the $4$ times heavier interactant.

The case of heavy envelopes can generate multiple peaks to the left of the original peaks due to multiple products.  According to the superposition law, the number of products scales with $A^N$ after $N$ interactions with envelope baryons of mass number $A$. So the later generations (tertiaries and so forth) whose energy are mostly below $10^{18}\,\rm eV$ are far more numerous than the earlier generations (primaries and secondaries). This brings the low end of the original peaks up to be a second, or even third additional peaks for all compositions; they also contribute to an increment of primaries around $4\times10^{19}\,\rm eV$ for helium envelopes and $1.3\times10^{19}\,\rm eV$ for carbon envelopes.

\section{Implications for the scenario of UHECR production in newly-born pulsars}\label{section:implications}

The success of a UHECR source scenario lies in its ability to reproduce these observations: i) the energy spectrum, ii) the composition, iii) the anisotropy, and iv) on the fact that it requires a rate of sources consistent with the population studies inferred from other astronomical observations. 

As we discuss in this section, the results obtained in this paper suggest that all four points could be reasonably achieved in the extragalactic rotation-powered pulsar scenario. Newly-born pulsars are natural candidates to reproduce points ii) and iii), due to their iron-peaked surface (if the composition at the highest energies proves to be actually heavy, as the measurements of Auger seem to indicate) and their transient nature. Though point i) is challenged by the fact that the toy model of unipolar induction generates a hard spectrum that does not fit the observed UHECR spectrum, our results show that the slope could be naturally softened during the escape from the supernova envelope (also seen in \citealp{BB04} for Galactic pulsars). The range of parameters for the pulsar and its surrounding supernova allowed for a successful acceleration and escape at the highest energies is relatively narrow. This potential issue is however compensated by two advantages. First, the range of values required for the initial parameters of both the pulsar and its supernova are close to the ones inferred for the youngest isolated pulsars observed nowadays (see e.g., Table~3 of \citealp{Chevalier05}). Second, the rate of such objects required to account for the observed flux of UHECRs is low, of order $f_{\rm s}\lesssim 0.01\%$ of the `normal' (as opposed to binary millisecond) pulsar birth rate. Point iv) can hence also be deemed as reasonably satisfied. 

We will also examine in what follows, the implications of our results on the arrival directions of UHECRs in the sky, and on possible probes of this source scenario. We also discuss the signatures expected for secondary messengers such as neutrinos, gamma-rays and gravitational waves. 

These implications are first discussed under the assumption that the currently observed UHECR flux has an extragalactic origin. The contribution of Galactic pulsar births is discussed in Section~\ref{subsection:Galactic}.

\subsection{Required source density and type of source}\label{subsection:flux}

The magnetar birth rate necessary to account for the observed flux of UHECRs was estimated in \cite{Arons03} and updated for various cases by \cite{K11}. The same calculations can be applied to our case, for rotation-powered pulsars. 

For a population of identical neutron stars with initial rotation velocity $\Omega_{\rm i}$ and magnetic dipole momentum $\mu$, satisfying Eq.~(\ref{eq:param_scan}), one can adapt the normalization found by \cite{K11} for negligible gravitational wave losses. An identical neutron star assumption is acceptable, in so far as the allowed parameter range of sources for particle acceleration and escape is fairly narrow (Eq.~\ref{eq:param_scan}). 
A birth rate of $\dot{n} \sim 10^{-8}\,\mu_{31}Z_{26}^{-1}\,$Mpc$^{-3}\,$yr$^{-1}$ is required to produce the observed UHECR flux, in the absence of source evolution history. When the emissivity of UHECR sources is assumed to follow the star formation history, the pulsar birth rate at $z=0$ is of order: $\dot{n}_{\rm SFR} \sim 0.8\, \dot{n} \sim 0.8\times 10^{-8}\,\mu_{31}\,$Mpc$^{-3}$~yr$^{-1}$.
This calculation assumes that the total Goldreich-Julian charge density is tapped in the wind for UHECR acceleration (Eq.~\ref{eq:spectrum_arons}). A lower efficiency would result in a lower energy flux per source, and thus in higher required densities. 

The above rates correspond to a fraction $f_{\rm s}\lesssim 0.01\%$ of the birth rate of `normal' pulsars, which is of order $1.6\times 10^{-4}$~Mpc$^{-3}$~yr$^{-1}$ (or one per 60 years in the Galaxy, which is consistent with the supernova rate, \citealp{Lorimer08}). Among the `normal' pulsar population, it is difficult to infer the number of objects that would satisfy Eq.~(\ref{eq:param_scan}), as the distribution of pulsars according to their initial rotation velocities and magnetic field is not straightforward (see examples of models discussed in \citealp{GL02} and \citealp{BB04}). 

\cite{Faucher06} find that the birth spin period distribution of pulsars is normal, centered at 300 ms and 
with standard deviation 150 ms, and that the initial magnetic field follows a log-normal distribution with $\langle \log (B/{\rm G}) \rangle\sim 12.65$ and $\sigma_{\log B}\sim 0.55$. They stress however that this distribution of birth spin periods is not precisely constrained by their method, and considerable deviations from this statistics could be expected. Such a distribution would imply that $\lesssim 2\%$ of the `normal' pulsar population could be endowed with sub-millisecond periods at birth.

Equation~(\ref{eq:param_scan}) further depend on the supernova characteristics (ejected mass and energy). However, as discussed in \cite{Chevalier05}, the pulsar properties do not appear to be closely related to the supernova category. This introduces an additional degeneracy on the type and total number of objects meeting the requirements for acceleration and escape. 
Nevertheless, it is promising that the range of values required for the initial parameters of both the pulsar and its supernova are close to the ones inferred for the youngest observed isolated pulsars (see e.g., Table~3 of \citealp{Chevalier05}).

Hence, $f_{\rm s}$ is a small enough fraction to leave reasonable room for poorer injection efficiencies, and to account for the narrowness of the parameter range of Eq.~(\ref{eq:param_scan}).

One should also keep in mind that both HiRes \citep{Abbasi09} and the Pierre Auger Observatory \citep{Abraham10} report systematic uncertainties of order 20\% on the absolute energy scale of the spectrum, which should be considered for the evaluation of $\dot{n}$.

The distribution inferred by \cite{Faucher06} implies that pulsars with birth periods $\sim 300 \pm 150 \,$ms are about $\epsilon\sim 30$ times more numerous than the submillisecond ones. Such pulsars could potentially accelerate iron up to $E (P=100\, {\rm ms}) \simeq 10^{16}\,$eV (Eq.~\ref{eq:Eacc}). For extragalactic pulsars with a similar acceleration mechanism to the case we discuss here (i.e., only a fraction $f_{\rm s}$ of the existing population leading to cosmic-ray production), the amplitude of the injected spectrum at these lower energies is well below the observed one (even if 30 times more numerous, the hard $\sim E^{-(1-1.5)}$ power-law, below the peak due to the secondary protons, only overtakes the observed $\sim E^{-3}$ spectrum closer to ankle energies).  At these low energies, the diffusion of cosmic-rays in the intergalactic magnetic fields would further prevent them from reaching us, if the sources are located at tens of megaparsec distances. On the other hand,  a {\it Galactic population} of these more numerous slower pulsar births may give important contributions to the cosmic ray spectrum below the ankle (see, e.g., \citealp{BB04}).

\subsection{Propagated escaped energy spectrum}\label{subsection:spectrum}

The cosmic ray spectrum observed by the Pierre Auger Observatory can be described as a broken power-law, $E^{-\lambda}$, with spectral index $\lambda\sim 3.3$ below the break (called ``ankle") around $10^{18.6}$~eV, and $\lambda \sim 2.6$ above, followed by a flux suppression above $\sim10^{19.5}$~eV \citep{Abraham10}.

One issue of the model advanced by \cite{Blasi00} and \cite{Arons03} for the acceleration of UHECRs in pulsars and magnetars is the hardness of the produced spectrum, that hardly fits the observations described above, even after propagation. These models were introduced in the ``AGASA era", to account for the absence of GZK cutoff in the observed spectrum \citep{Takeda98}. They aimed at producing a hard spectrum (of spectral index $-1$, see Eq.~\ref{eq:spectrum_blasi}) to fit the highest energy end of the spectrum, beyond $E>6\times 10^{19}\,$eV, and do not fit the slope at lower energies. The latest experiments report however that a suppression reminiscent of the GZK cut-off is present at the highest energy end of the UHECR spectrum \citep{Abbasi08,Abraham10}. Hence, a hard spectrum need no longer be advocated to explain the measurements, and now constitutes a disadvantage.

\cite{K11} proposed to alleviate this issue by introducing a distribution of initial parameters of magnetars among their population (see also \cite{GL02} for the Galactic pulsars case). Such a distribution results in a distribution of the maximum acceleration energy, and adequate values can be found to soften the integrated spectrum and fit the observations. The same calculation can be applied to the case of rotation-powered pulsars. 

Note also that in order to have the monoenergetic-type acceleration spectrum given in Eq.~(\ref{eq:spectrum_arons}), the wake-field acceleration which is based on the ponderomotive force requires the magnetic field to be coherent over the acceleration region. However, much smaller coherence scales can be naturally expected, leading to a stochastic acceleration, that could also produce a $E^{-2}$ spectrum. Such cases have been studied in different contexts by e.g., {\cite{Chen02,Chang09}}.

The results of \cite{BB04} already show that the injection of iron nuclei and their escape through the pulsar nebula can lead to a softer spectrum due to the production of secondary nuclei. This feature was however not deeply discussed and highlighted, as the AGASA energy spectrum available at that time highly differed from the current observations. Besides, the calculations of \cite{BB04} are based on simplified hadronic interaction cross-sections, and on the raw assumption that one interaction leads to the fragmentation of the primary nucleus in two nuclei with different mass numbers. 

Our detailed analysis demonstrates that, within the range of pulsar and supernova envelope parameters given in Eq.~(\ref{eq:param_scan}) and Fig.~\ref{fig:contours}, the injection of heavy nuclei and their escape from the envelope naturally enables the softening of the energy spectrum to indices of order $\sim 1.5-2$ (Figs.~\ref{fig:iron}, \ref{fig:real}). 
As explained in Section~\ref{subsection:numerical}, this softening stems from the abundant production of secondary nucleons, helium and intermediate nuclei at low energies.

After propagation and interactions in the intergalactic medium, the injection of particles at the source with index $\sim 2$ is expected to provide a good fit to the observed UHECR spectrum. Our escaped composition can be identified with the mixed composition introduced by \cite{Allard08} (see also \citealp{Aloisio11}) that contains 30\% of iron and assumes a maximum proton energy of $E_{p, {\rm max}}\sim10^{19}\,$eV. \cite{Allard08} calculates that an injection index of order $2.0-2.1$ is required to adjust the observed UHECR spectrum after propagation through the intergalactic medium. If one assumes that the source emissivity in UHECRs has evolved according to the star formation rate, the required injection index at the source is of $\sim 1.2$ \citep{KAO10}.

The bumps and irregularities apparent in the escaped spectra (Figs.~\ref{fig:iron}, \ref{fig:real}) 
should be attenuated by the propagation, a possible distribution of neutron star characteristics (essentially a distribution of the dipole moment $\mu$, initial spin $\Omega_{\rm i}$), and especially the envelope chemical composition. 

The flux of particles with energy below the ankle should not overwhelm other (possibly Galactic) components. Our calculations show indeed that the escaped spectrum should become harder below $E\sim 10^{18}\,$eV, with a slope of order $-1.5$ due to the tail of secondary protons. The flux of these lower energy particles should also be diluted by the large dispersion of their arrival times, after propagation in the intergalactic and Galactic magnetic fields.

The injection of a pure proton composition by neutron stars is likely only viable in models where the envelope column density is thinner by many orders of magnitudes compared to classical 
supernov{\ae}  at early times. 
In this situation, the resulting UHECR observable quantities are similar to what has been discussed until now: a hard spectrum injection should be expected after escape (spectral index $-1$), that could be reconciled with the observed spectrum by invoking a distribution of neutron star characteristics, as in \cite{K11}.\\

One probe of this scenario (both in the proton or iron-rich injection cases) would be a sharp cut-off of the energy spectrum at energies above $E_{\rm cut,Fe}$ (or $E_{\rm cut,p}$ for pure proton injection). A mild recovery is indeed expected if the maximum acceleration energy were $E>10^{20.5}\,$eV, as the observed cut-off in the spectrum would then be due to the GZK effect.

\subsection{UHECR composition}

Recent measurements by the Pierre Auger Observatory indicate that the cosmic ray composition transitions from being dominated by protons below the ankle ($\sim 10^{18.6}$~eV) to being dominated by heavier nuclei with average masses similar to Si or Fe at $\sim 10^{19}\rm eV$ \citep{Abraham:2010yv}. The instruments located in the Northern hemisphere, HiRes and Telescope Array, seem to observe a light composition up to the highest energies, though the results of the former remain consistent with those of Auger within errors.{} We caution furthermore that the composition measured by these experiments concern energy bins below $\sim 4\times 10^{19}\,$eV, due to the lack of statistics at the highest energy end.

Neutron stars are one of the most likely places to inject heavy nuclei abundantly, as we discussed in Section~\ref{subsection:injection}. It is interesting to notice that the escaped composition resulting from such an injection indicates a transition from light to heavy nuclei around the energy observed by Auger (Figs.~\ref{fig:iron}, \ref{fig:real}). 

Note that \cite{BB04} found a similar transition, but did not devote much discussion on that feature. That finding was not necessarily appealing during the AGASA era, when the composition was believed to be light at the highest energies. To account for the continuation of the flux above GZK energies, \cite{BB04} added to their Galactic pulsar population, a pure proton extragalactic component, which lightens their overall composition at the highest energies. 

As mentioned in the previous section, our escaped composition is similar to the low $E_{p, {\rm max}}$ mixed composition introduced by \cite{Allard08} and \cite{Aloisio11}. The resulting composition after propagation in the intergalactic medium when such a composition is injected is shown to conserve the transition between light to heavy elements around $10^{19}\,$eV \citep{Allard08}.

The injection of a mixed composition with $\lesssim 10\%$ of iron would remain consistent with such a transition. Indeed, Figs.~\ref{fig:iron}, \ref{fig:real} show that the rate of secondary protons is more than 10 times higher than the rate of injected iron. Injected protons would cut-off below $E_{\rm cut,p}$ and would not overwhelm the escaped iron flux at the highest energies. 

One can note that, depending on the detailed transition from extragalactic to Galactic component, the   composition found here may induce an anisotropy signal at lower energies as was discussed in \cite{LW09}.
With such a dominant heavy composition at ultrahigh energies ($\gtrsim 10^{19.7}\,$eV), one expects that any anisotropy signal at the highest energies would have a similar structure around $2\,$EeV where the composition is proton dominated, about two times stronger. Such an anisotropy is not observed by the Auger observatory \citep{Abreu11}, which may question the composition of the mild anisotropy found at the highest energies or imply a more complex composition structure both for the extragalactic as well as the Galactic component around EeV.

The injection of a pure proton composition is not ruled out either in our scenario, but is only favored under stringent conditions on the early envelope density. 

\subsection{Distribution of events in the sky}

The radio, X-ray and gamma-ray signals of rotation-powered pulsars and magnetars are too weak to allow their detection beyond our Local Group. For this reason, a direct spatial coincidence between a neutron star and UHECR arrival directions is not expected to be observed, if the source is not born inside our Local Group.

Nevertheless, the distribution of UHECR events could follow the large scale structures, where neutron stars should be concentrated. In particular, these objects should be frequently found in star forming galaxies. Such distributions would be apparent only if the deflections experienced by particles in the Galactic and intergalactic magnetic fields are small. Moreover, anisotropic signatures would only be distinguishable for ensembles of particles with the highest energies. Above $E\sim E_{\rm GZK}\equiv 6\times10^{19}\,$eV, the horizon that particles can travel without losing their energy is limited to a few hundreds of megaparsecs and the distribution of sources in that local Universe appears anisotropic.

Neutron stars can be considered as transient UHECR sources. Cosmic rays with energy above $E_{\rm GZK}$ can indeed only be produced during the first $\Delta t_{\rm s}  \sim 4\,Z_{26}\eta_1I_{45}\mu_{30.5}^{-1}\,$yr after the birth of the neutron star. 

This implies that if secondary messengers such as neutrinos, gamma-rays, or gravitational waves were produced at the same time as UHECRs, they would not be observed in temporal coincidence with the latter. The time delay experienced by UHECRs in the intergalactic magnetic field is indeed of order $\sim 10^4$\,yrs for one degree deflection over 100~Mpc, which is much longer than the duration of the UHECR production.

Transient sources could lead to bursts of events in the sky if the number of event per source is important, and the arrival times of particles {\it is not diluted by the dispersion induced by magnetic deflections} \citep{Kalli10}. For extremely high energy protons and low intergalatic fields,  the total dispersion time due to magnetic deflection, $\Sigma_t$, can be shorter than the detector exposure time, $T_{\rm exp}$, the number of sources contributing to the observed UHECR flux inside a radius of $l=100\,$Mpc is of order $N_{\rm s}=(4\pi/3)l^3\dot{n} T_{\rm exp}\sim 0.4$, using the pulsar birth rate inferred in the previous section, and $T_{\rm exp}=10\,$yrs.  The number of events that can be detected from each source is: $N_{\rm ev} = {\cal E}_{\rm UHECR}A_{\rm exp}/(E_{\rm GZK}\, 4\pi l^2)\sim 2\times 10^3$, where we assumed a detector exposure of $A_{\rm exp}=3000\,$km$^2$, as for the Pierre Auger Observatory, and the cosmic-ray energy output per source ${\cal E}_{\rm UHECR}\sim 5\times10^{51}\,$erg in our milli-second pulsar scenario. It is likely however that cosmic rays arriving from most directions in the sky experience a significant dispersion in their arrival time, due to magnetic fields: $\Sigma_{\rm t}>T_{\rm exp}$. This should be the case for iron nuclei, unless one assumes unrealistically low magnetic fields. In that case, the number of events detected from one source would be reduced by a factor $T_{\rm exp}/\Sigma_t$. The reader can refer to \cite{KL08b} and \cite{Kalli10} for detailed discussions on the dependence of $\Sigma_t$  on magnetic field parameters. 

A direct identification of the source could be possible if a pulsar was born inside our Galaxy, or close enough to allow X-ray or gamma-ray observations.
The dispersion of arrival times inside our Galaxy $\sigma_{\rm Gal}$ reads:
\begin{eqnarray}\label{eq:tGal}
\sigma_{\rm Gal} &\sim& 2.5\,Z^{2}\left( \frac{l}{10\,{\rm kpc}}\right)^2\left( \frac{B_{\rm turb}}{4\,\mu{\rm G}}\right)^2\times\nonumber\\&&\left( \frac{\lambda_{\rm turb}}{50\,{\rm pc}}\right) \left( \frac{E}{E_{\rm GZK}}\right)^{-2}\,{\rm yr} .
\end{eqnarray}
Here we noted $B_{\rm turb}$ and $\lambda_{\rm turb}$ the turbulent magnetic field intensity and coherence length respectively, and $l$ the distance of the source. The time delay  $\delta t_{\rm Gal}$ experienced by particles due to the turbulent Galactic magnetic field is typically much larger than $\sigma_{\rm Gal}$, due to the additional deflection due to the regular magnetic field component. 

For proton injection and a weak regular magnetic field component, this implies that if such an event were to occur, a burst in UHECRs with a typical rise and decay timescale of a fraction of year would be observed in the sky, from a time $\delta t_{\rm Gal}$ after the onset of the explosion that triggered the birth of the fast-spinning neutron star. In this case, secondary messengers propagating rectilinearly would also arrive at a time $\delta t_{\rm Gal}$ before UHECRs.

For iron nuclei injection, the highest energy elements come out of the envelope as heavy nuclei. These should reach the Earth after a time delay of $\delta t_{\rm Gal}\gtrsim 1750\,$yrs$\,/l_{10\,\rm kpc}^2$. For very close-by sources (e.g., at 2 kpc), $\delta t_{\rm Gal}$ could be of order $T_{\rm exp}$, leading to a sudden increase in the detection of ultrahigh energy events (about $N_{\rm ev}\sim 4\times 10^{12}$ over $\delta t_{\rm Gal}\gtrsim 70\,$yrs, for a source at 2\,kpc). In that case, the time order of escape of the different chemical elements from the envelope should be washed out by the fact that $\delta t_{\rm Gal}>\Delta t_{\rm s}$.

Particles at energies $E < E_{\rm GZK}$ should arrive with more consequent time delays, so potentially from young rotation-powered pulsars that are detected nowadays. The dispersion in time should however be as consequent, and such events should not be detected as bursts, but only as continuously arriving particles. No spatial clustering from such sources is expected either, as the deflections experienced by particles at these low energies should be large. 

If EeV or higher energy neutrons were produced by these objects, by interactions of accelerated nuclei in the envelope for example, they would propagate rectilinearly and would appear as point sources. However, nearly no time delay between the detection of the birth of the neutron star and the arrival of the particles is expected. Spatial correlations between pulsar positions and neutron events are thus expected only if a new birth actually occurs in the Galaxy.

\subsection{Secondary messengers}

The propagation of UHECRs in the intergalactic medium should lead to the production of cosmogenic neutrinos and gamma-rays by interactions on the Cosmic Microwave Background. The expected cosmogenic neutrino and gamma-ray fluxes depend mostly on parameters inherent to cosmic-rays themselves (their composition and overall flux), but also on the injection index at the source and the source emissivity evolution history for diffuse fluxes (see e.g., \citealp{KAO10} for a parameter scan over these astrophysical variables). The cosmogenic gamma-ray signatures further depend on the structure and strength of the intergalactic magnetic fields, because of the pair production/inverse Compton cascading of photons in the intergalactic medium.

For a source evolution following the star formation rate, as can be expected for neutron stars, an injection of pure proton or proton-dominated compositions with power-law spectral index $\sim 2.0-2.5$ would successfully fit the observed UHECR spectrum. The resulting diffuse cosmogenic neutrino flux would lie within the gray shaded region of Fig.~9 of \cite{KAO10}. For an iron dominated injection up to a few times $10^{20}\,$eV and a proton dominated injection below $10^{19}\,$eV (as we get in Figs.~\ref{fig:iron}, \ref{fig:real}) one expects a lower neutrino flux, peaking around $E_{\nu}\sim10^{8.5}\,$GeV with $E_\nu^2({\rm d}N/{\rm d}E_\nu)|_{\rm max}\sim 5\times 10^9\,$GeV~cm$^{-2}$~s$^{-1}$~sr$^{-1}$ (red dash-dotted line of Fig.~7 of \citealp{KAO10}). For the diffuse cosmogenic gamma-ray background, the same fit to the observed UHECR spectrum gives fluxes peaking around $E_\gamma\sim10\,$GeV of order $E_\gamma^2({\rm d}N/{\rm d}E_\gamma)|_{\rm max}\sim7\times 10^{-13}-10^{-12}\,$eV\,m$^{-2}$\,s$^{-1}$~sr$^{-1}$ for both proton dominated compositions and for our proton to iron transition scenario (see Figs.~4 and 8 of \citealp{Decerprit11}).

For single sources, \cite{Decerprit11} showed that the cosmogenic neutrino flux could be within reach of IceCube for powerful steady sources {(see also \citealp{Essey2010})}. Only beamed sources (i.e., blazars) seem to satisfy the required luminosity condition (otherwise, the required power exceeds the Eddington power), but the neutrino flux is then diluted by the deflection of cosmic rays  { \citep{Murase11}}.  In the case of transient sources, the total received flux should be diluted by the ratio of the emission time to the spread in the arrival times due to the magnetic fields, $\Delta t_{\rm s}/\Sigma_t$, which could lower the flux of many orders of magnitude, preventing any detection. In the same token, as was discussed in \cite{GA05} and \cite{KAL11}, the secondary gamma-ray emission (produced in the intergalactic medium) from a single transient source should be affected by dilution in time, and be below reach of next generation gamma-ray instruments.

\cite{Murase09} calculated that a promising amount of neutrinos would also be produced via hadronic interactions during the escape of ultrahigh energy protons from the surrounding supernova envelope. These authors show the importance of muon and pion interactions on the baryonic envelope for the out-coming neutrino flux, especially for interactions at the earliest times.

Our current simulations do not take into account such interactions, hence an accurate evaluation of the neutrino flux associated to our scenario cannot be computed with the present tool. It can be noted however, that the overall background neutrino flux that we would obtain would be similar to the neutrino flux predicted by \cite{Murase09} for proton injection and at least about one order of magnitude smaller for iron injection. Our lower required source rate should not affect the neutrino flux, as its level is determined mainly by the energy injected in UHECRs above the pion production threshold energy. The supernova envelope opacity necessary to allow the escape of iron nuclei at the highest energies results indeed in a neutrino flux about one order of magnitude lower than in the case of protons \citep{Murase10}. We also calculated that the primary injection of iron would lead to an enhanced flux (of a factor of a few) around PeV energies because of the steeper overall UHECR spectrum generated after the escape. The level of neutrino flux could still turn out to be fairly high (the flux predicted by \citealp{Murase09} is significantly above the IceCube sensitivity), and leaves room for detection with the IceCube experiment, either of single close-by sources born within $\sim 5\,$Mpc, or of the diffuse background. A full calculation of the expected flux is needed to formulate more quantitative statements. 

The gamma-rays produced in the supernova envelope via hadronic interactions  could cascade in turn on the same background and escape as photons in the TeV range. This process could possibly produce a bright transient gamma-ray source, though the exact spectrum and its detectability have to be quantitatively calculated. A fraction of ultrahigh energy photons could also escape, that could be also observed as a transient source by experiments such as Auger or JEM-EUSO, for sources at a few megaparsec distances { \citep{Murase09b}}. Again, these assertions need more careful investigations.

Highly magnetized magnetars with fields $B \gtrsim 10^{15}\,$G should be strong emitters of gravitational waves. If protons are injected by pulsars, the hard produced spectrum requires a specific distribution of pulsar parameters (of their initial rotation velocity and/or their magnetic field strength) to soften the overall UHECR spectrum and fit the observations \citep{K11}. In such a case, and for strong pulsar internal deformations, \cite{K11} argues that a characteristic diffuse gravitational wave signal would be produced, that could be detected with future generation detectors such as DECIGO or BBO. The present study shows however that these strong magnetic fields would induce a fast spin-down time that could not allow the escape of UHECRs in presence of a dense supernova shell. The problem could be bypassed if the envelope is  particularly under-dense, if particles could escape through a breech created by a proto-pulsar jet (but interactions with the radiative background would no longer be negligible in that case), or for envelope shredding scenarios as invoked by \cite{Arons03}.
For the milder fields favored in our scenario ($B\sim 10^{13}\,$G), the gravitational wave signal is expected to be lower by many orders of magnitude, far below the reach of any planned instruments.

\subsection{A Galactic scenario for UHECRs?} \label{subsection:Galactic}

We discuss in this section the scenario in which the major contributor to the currently observed UHECR flux are Galactic pulsars injecting iron, and possessing the parameters required for iron escape at the highest energies. The pure proton injection case seems indeed difficult to reconcile with a continuous detection of UHECR events, given the short spread in their arrival times $\delta t_{\rm Gal}$ at the highest energies and the lack of anisotropies toward the Galactic plane. The iron injection case could be more promising, in so far as $\delta t_{\rm Gal}$ can be much longer than the detector exposure time, $T_{\rm exp}$, even at the highest energies, for reasonable Galactic magnetic field strengths. 

We note $\nu_{\rm s}$ the birth rate of neutron stars in our Galaxy that satisfy the conditions for successful iron acceleration and escape at the highest energies (Eq.~\ref{eq:param_scan}). We recall that the `normal' pulsar birth rate is of order $\nu_{\rm Gal}\sim 1/60\,$yr$^{-1}$ in the Galaxy, \citep{Lorimer08}. If the time interval between two births is shorter than the dispersion of the arrival times $\nu_{\rm s}^{-1}<\delta t_{\rm Gal}$, then the flux of UHECRs should not depend on $\delta t_{\rm Gal}$ and could be accounted for by the fraction $\sim10^{-7}-10^{-6}$ of the population of pulsars within our Galaxy.

Now, if $\nu_{\rm s}^{-1}>\delta t_{\rm Gal}$, one may have zero (in which case the Galactic scenario does not stand) or only one source  contributing to the observed Galactic UHECR flux. The UHECR flux due to this source can be written:
\begin{eqnarray}
E^3J_{\rm 1s} (E)& = & E^3\frac{{\rm d}N_{\rm i}}{{\rm d} E}\frac{1}{(4\pi)^2l^2}\frac{1}{\delta t_{\rm Gal}}\\
&\sim& 4\times10^{30}\,{\rm eV}^2\,{\rm m}^{-2}\,{\rm s}^{-1}\,{\rm sr}^{-1}\frac{I_{45}}{Z_{26}\mu_{30.5}l_{10\,{\rm kpc}}^{2}}\nonumber\\
&&\times\left(  \frac{E}{E_{\rm GZK}}\right)^2 \left(\frac{\delta t_{\rm Gal}}{2\times 10^3\,{\rm yr}}\right)^{-1},
\end{eqnarray}
This estimate implies that in this scenario, the cosmic-ray injection efficiency should be $\sim 4\times10^{-6}$ times lower not to overshoot the spectrum. 

In the scenario where extragalactic pulsars dominate the observed UHECR flux, $\xi\nu_{\rm s}=f_{\rm s}\nu_{\rm Gal}$, with $\xi$ the iron injection efficiency. With $f_{\rm s}\lesssim 0.01\%$ as calculated before, we would likely fall in the latter case, with $\nu_{\rm s}^{-1}>\delta t_{\rm Gal}$. In this scenario, there should be no Galactic source contributing currently, as otherwise, it would overshoot the observed spectrum. 

Note also that these flux estimates are subject to strong variations according to the structure and strength of the Galactic magnetic field.

In both cases (single or many Galactic sources contributing), the energy spectrum should present a cut-off (at $E_{\rm cut}$) mimicking the GZK cut-off, due to the propagation of particles in the supernova envelope. The ankle feature would stem from the change in slope around the secondary proton peak. The overall spectral index could fit the observed one by a combination of the escaped spectrum and the propagation effects in the Galaxy. 

The chemical composition of UHECRs detected on Earth would slightly differ from the composition of particles escaped from the supernova envelope. Protons would indeed disappear more quickly from the Galaxy than heavy elements around $10^{19}\,$eV, as $\delta t_{\rm Gal}$ of order of $T_{\rm exp}$ at this energy. At lower energies, particles should still be able to remain confined, and a transition from light to heavy nuclei should still occur.

Finally, the main weakness of this Galactic scenario lies in the expected anisotropy signature. A single source should lead to a noticeable spot of events in the sky at the highest energies, even for iron nuclei, unless the turbulent Galactic magnetic field is extremely strong. If many sources were contributing, they are also expected to trace the Galactic disk, but no such anisotropy has been observed in the UHECR data.

\section{Conclusions}\label{section:conclusion}
We studied the injection and escape of UHECRs from newly-born pulsars based on a Monte Carlo simulation of hadronic interactions and on a detailed examination of the physical properties of supernov\ae{} envelopes. 
Our results show that protons and light elements at the highest energies can traverse the envelope only for very dilute envelopes. For pulsars embedded in supernov\ae{} with characteristics satisfying Eq.~(\ref{eq:param_scan}), iron nuclei are able to escape from the supernova envelope with energy above $10^{20}\,\rm eV$. The escaped spectrum displays a transition from light to heavy composition at a few EeV, matching the recent Auger data. Due to the production of secondary nucleons, the escaped spectrum also presents a softer slope than the initial injected one, enabling a good fit to the observations. 

Under the assumption that unipolar induction acceleration can take place in the neutron star winds, 
two conditions ensure the compelling adequacy of the scenario of production of UHECR by neutron stars with the observed data: 

\begin{enumerate}
\item that a fraction $f_{\rm s}\lesssim 0.01\%$ of extragalactic supernov\ae{} give birth to pulsars with sub-millisecond periods, and dipole magnetic field in the range $10^{12-13}\,$G, 

\item that a successful injection of heavy nuclei is occurring at the acceleration site of these objects.
\end{enumerate}

We discussed that this double condition can be reasonably fulfilled.  Indeed, about 2\% of young `normal' pulsars are inferred to have initial parameters close to fulfilling condition 1. The low value of $f_{\rm s}$ also permits poorer injection efficiencies and compensates for the narrowness of the allowed parameter range. Condition 2 is naturally favored in neutron stars that have heavy nuclei rich surfaces. Should these two conditions be fulfilled, the main UHECR observables, namely, the energy spectra, composition, and arrival directions, would be consistent with the latest Auger data \citep{Abraham10, Abraham:2010yv,Abreu10}. 

If criterion 2 is not met, and only protons or light elements can be accelerated, the neutron star scenario is viable only for dilute surrounding envelopes, or if mechanisms shredding or piercing the envelope are at play to enable the escape of particles. 

In the iron injection case, the birth of such an object within our Galaxy would be noticeable in the number of detected events, only for very close-by sources (at $\sim 2\,$kpc). Such a source could lead to a distinct increase of the observed UHECR events starting $\delta t_{\rm Gal}\gtrsim 70\,$yrs $\times(l/2\,{\rm kpc})$ after the birth (for a source located at $l$, with parameters chosen in Eq.~\ref{eq:tGal}), and that would last for $\delta t_{\rm Gal}$. If a pulsar birth were observed today, the proton injection case and/or the production of neutrons by interactions in the direct environment of the source would lead to a significant burst of UHECR events in the sky, a fraction of year later. 
The birth rate of neutron stars satisfying our criteria inside our Galaxy is however expected to be as low as $\sim 5\times 10^{-7}\,$yr$^{-1}$. 

Other signatures can be expected, such as a non-recovery of the energy spectrum above $E_{\rm cut,Fe}\sim 10^{20.5}\,$eV, or the precise measurement of the cosmic ray composition at high energies. Large exposure instruments such as Auger North or JEM-EUSO would allow to make such measurements and probe this scenario.

\acknowledgments
We thank V. Dwarkadas, C. Fryer, S. Horiuchi, M. Lemoine, H. Li, B. Metzger, K. Murase, S. Phinney, T. Pierog, and the Auger group at the University of Chicago for very fruitful discussions. This work was supported by the NSF grant PHY-1068696 at  the University of Chicago, and the Kavli Institute for Cosmological Physics through grant NSF PHY-1125897 and an endowment from the Kavli Foundation.

\appendix

\section{Alternative mechanisms to escape the supernova envelope?}\label{app:alternative}

Recent works have shown the possibility that the confining pressure of the toroidal magnetic field could collimate the proto-magnetar wind along its polar axis, and drive a jet that has the properties of long gamma-ray bursts jets \citep{Komissarov07,Bucciantini07,Bucciantini08,Bucciantini09}. This scenario opens up the possibility that cosmic rays be accelerated via magnetic reconnection or Fermi acceleration inside the proto-magnetar jet, and escape through the pierced supernova envelope.
The case of nuclei escaping through a jet has been discussed semi-analytically in the context of GRBs by \cite{Murase08} and for proto-magnetar jets by \cite{Metzger11}. 

However, mildly magnetized pulsars could not have the collimation power to produce a jet.
As discussed in \cite{Bucciantini07}, the collimation becomes significant for  values of the ratio of the Poynting flux to the total energy at the termination shock of the wind, $\dot E_{\rm mag}/\dot E_{\rm tot} \gtrsim 0.2$, at times $t\sim 10-100$~s. The conversion of magnetic energy into kinetic energy in relativistic outflows at large radii is uncertain. For the mild magnetic fields and high rotation velocities that we consider, the magnetization at the light cylinder reads $\sigma_{\rm L}\equiv 4\mu^2\Omega^4 /(\dot M c^5)\sim 20 \mu_{30.5}^2\Omega_4^4/\dot M_{9.5}$, where the mass loss rate at $t\sim 30$~s, $\dot M_{9.5}\equiv \dot M/10^{-9.5}\,M_\odot$, was inferred from Appendix A2 of \cite{Metzger10}. The value of $\sigma_{\rm L}$ should increase steeply with the fast mass loss rate around $t\sim 60$~s and later (when the wind becomes transparent to neutrinos). For such high $\sigma_{\rm L}\gg 1$, it is plausible that magnetic dissipation in the relativistic outflow out to the termination shock leads to a low $\dot E_{\rm mag}/\dot E_{\rm tot}$ at these distances \citep{Coroniti90,Lyubarsky01,Kirk03}, not allowing the formation of a jet. Studies of the Crab Pulsar wind nebula show indeed that $\dot E_{\rm mag}/\dot E_{\rm tot}\sim 10^{-2}$ at large radii \citep{Kennel84,Begelman92}.

One may note that, even in a scenario where a jet were produced (for more strongly magnetized neutron stars), our conclusions would still apply, for particles that would not be injected in the direction of the jet, but in the other sectors. Such particles would have to cross the expanding supernova shell, and would have the same fate as in the present framework.

Another mechanism to bypass the problem of UHECRs crossing the dense supernova envelope was invoked by \cite{Arons03},
who proposed that the supernova envelope be disrupted by the magnetar wind. Such phenomena have never been observed, neither in magnetar envelopes, nor in rotation-powered pulsar envelopes. 

Finally it is also possible that millisecond pulsars can be born in Accretion-Induced Collapse of white dwarfs \citep{Fryer99, Fryer09}. The escape scenario is different and will be studied in our future work.

\bibliography{FKO11}

\end{document}